\documentclass[aps,prd,notitlepage,nofootinbib,superscriptaddress]{revtex4-1}
\usepackage[utf8]{inputenc}
\usepackage[table ]{ xcolor}

\usepackage{multirow}

\usepackage{hyperref}
\usepackage{graphicx}
\usepackage{mathtools}
\usepackage{amsmath,amssymb,amsfonts,amsthm,latexsym,stmaryrd,physics,mathrsfs}
\numberwithin{figure}{section}
\allowdisplaybreaks[4]
\usepackage{marginnote}
\usepackage{color}
\usepackage{bm}
\usepackage{float}
\usepackage{nicefrac}
\usepackage{microtype} 
\usepackage{booktabs}
\usepackage{verbatim}
\usepackage{setspace}
\usepackage[T1]{fontenc}
\usepackage[utf8]{inputenc}
\usepackage{stfloats}
\usepackage{diagbox}
\usepackage{dsfont}
\usepackage{tikz}
\usepackage{geometry}
\usepackage{xtab}
\usepackage{subcaption}

\usepackage[utf8]{inputenc}
\usepackage[T1]{fontenc}
\usepackage{amsmath}
\usepackage{amsfonts}
\usepackage{amssymb}
\usepackage[version=4]{mhchem}
\usepackage{stmaryrd}
\usepackage{hyperref}
\hypersetup{colorlinks=true, linkcolor=blue, filecolor=magenta, urlcolor=cyan,}
\usepackage{bbold}
\usepackage{graphicx}
\usepackage[export]{adjustbox}
\graphicspath{ {./images/} }
\usepackage{mathrsfs}

\hypersetup{
	colorlinks=true
,urlcolor=blue
,anchorcolor=blue
,citecolor=blue
,filecolor=blue
,linkcolor=blue
,menucolor=blue
,pagecolor=blue
,linktocpage=true
,pdfproducer=medialab
}

\def\beq{\begin{equation}}
\def\eeq{\end{equation}}
\newcommand{\bea}{\begin{eqnarray}}
\newcommand{\eea}{\end{eqnarray}}
\def\bi{\begin{itemize}}
\def\ei{\end{itemize}}
\def\ba{\begin{array}}
\def\ea{\end{array}}
\def\bfig{\begin{figure}}
\def\efig{\end{figure}}

\newtheorem{theorem}{Theorem}[section]

\newcommand{\Slc}{\mathrm{SL}(2,\mathbb{C})}

\def\be{\begin{eqnarray}}
\def\ee{\end{eqnarray}}

\newcommand{\ck}{\mathcal K}

\newcommand{\sig}{\sigma}

\renewcommand{\l}{\lambda}

\renewcommand{\O}{\Omega}

\usepackage{geometry}
\begin{document}

\title{Properties of 4D spinfoam quantum geometry: Results from next-to-leading order spinfoam large-$j$ asymptotics of 1-5 Pachner move}

\author{Haida Li}
\email{eqwaplay@scut.edu.cn}
\affiliation{School of Physics and Optoelectronics, South China University of Technology, Guangzhou 510641, China}

\author{Muxin Han}
\email{hanm@fau.edu}
\affiliation{Department of Physics, Florida Atlantic University, 777 Glades Road, Boca Raton, FL 33431-0991, USA}
\affiliation{Department Physik, Institut für Quantengravitation, Theoretische Physik III, Friedrich-Alexander Universität Erlangen-Nürnberg, Staudtstr. 7/B2, 91058 Erlangen, Germany}

\author{Hongguang Liu}
\email{liuhongguang@westlake.edu.cn}
\affiliation{Institute for Theoretical Sciences, Westlake University, Hangzhou 310024, Zhejiang, China}
\affiliation{Department Physik, Institut für Quantengravitation, Theoretische Physik III, Friedrich-Alexander Universität Erlangen-Nürnberg, Staudtstr. 7/B2, 91058 Erlangen, Germany}

\author{Shicong Song}
\email{ssong2019@fau.edu}
\affiliation{Department of Physics, Florida Atlantic University, 777 Glades Road, Boca Raton, FL 33431-0991, USA}

\author{Dongxue Qu} 
\email{Corresponding author: dqu@perimeterinstitute.ca}
\affiliation{Perimeter Institute for Theoretical Physics, 31 Caroline St N, N2L 2Y5 Waterloo, ON, Canada}

\begin{abstract}


This paper proposes several criteria to probe the non-trivialities of 4-dimensional geometry that impact spinfoam amplitude. These criteria include the standard deviation of 4-volumes of the constituting 4-simplices, the smallest 4-simplex volume, and whether the directions of tetrahedron 4-normals are close to the null direction. By numerically computing and analyzing the spinfoam amplitudes up to the next-to-leading order of 1-5 Pachner move samples with the same boundary 4-simplex, we reveal the relationship between 4-dimensional geometry and spinfoam amplitudes, as large standard deviation of 4-simplex volumes and small 4-simplex volume can result in both leading order and next-to-leading order amplitudes being large. Furthermore, the numerical result indicates that the distance from tetrahedron 4-normals to the null direction has a greater impact on increasing the next-to-leading order amplitude than on the leading order amplitude, making it primarily a quantum effect.

\end{abstract}

\maketitle

\tableofcontents

\section{Overview}
Loop Quantum Gravity (LQG) is a background-independent approach to quantum gravity \cite{Ashtekar:2004eh,Han:2005km,thiemann2008modern}. The spinfoam model \cite{Perez:2012wv,rovelli2015covariant} is a covariant approach to LQG. One of the most successful spinfoam models currently is the Lorentzian Engle-Pereira-Rovelli-Livine (EPRL) model \cite{Engle:2007wy}, where the weak simplicity constraint is imposed \cite{Ding:2009jq,Ding:2010ye,Ding:2010fw}. The semi-classical behavior of spinfoam formalism is understood in terms of the large-$j$ (spin) asymptotics \cite{Barrett:2009mw,Rovelli:2010vv,Han:2011re1}. This asymptotic behavior is related to the Regge action of the classical discrete gravity  \cite{Conrady:2008ea,Han:2013gna,Liu:2018gfc,Han:2021rjo,Han:2021bln}. Recently, large-$j$ asymptotics of spinfoam EPRL models have been studied using numerical methods such as the stationary phase approximation \cite{Han:2020fil,Han:2021kll,Dona:2022yyn,Han:2023cen} and the method of Markov Chain Monte-Carlo on the Leftschetz thimble \cite{Han:2020npv,Dona:2022yyn}. Also, there's recent progress on the Regge calculus of asymptotic behavior of 1-5 move, which may also shed some light on the 1-5 move spinfoam large-$j$ asymptotics \cite{Borissova:2024txs}.

In this work, we focus on the spinfoam amplitudes of the 1-5 Pachner move complex. The 1-5 Pachner move operates on a configuration of a single $4$-simplex by adding a vertex in its bulk to obtain the final configuration of five 4-simplices sharing an internal vertex. It is interesting to study the spinfoam amplitudes of the 1-5 Pachner move primarily for two reasons: Firstly, since the 1-5 move can be constructed directly from a single 4-simplex as its boundary, which can be viewed as a step of fine-graining of the EPRL spinfoam model. Secondly, spinfoam amplitudes of flat geometries with non-trivial vertices in the bulk region are technically challenging to compute due to the presence of an infinite number of critical points, all leading to the same flat boundary geometry. As one of the simplest geometrical configurations of this type, the 1-5 Pachner move can serve as a suitable example for investigating such an issue.

In order to fully compute the spinfoam amplitude of the 1-5 Pachner move, an overall integration of the partition function is required. In the cases studied in previous works \cite{Han:2020fil,Han:2021kll,Han:2023cen}, this integration can be computed using the saddle point approximation. However, for the case of the 1-5 Pachner move, due to the presence of continuous critical points corresponding to all possible locations of the internal vertex, this integration can no longer be directly computed. Instead, we first need to extract the area parameters of four internal faces, which correspond to the degenerate directions of the Hessian matrix. The position of the internal vertex is fixed after the extraction. Then, the integration over all other degrees of freedom for each fixed internal vertex can be performed numerically using the saddle point approximation. In theory, it is possible to compute the total spinfoam amplitude numerically using Monte Carlo integration. However, in practice, this computation can be extremely difficult for two reasons: First, computing each individual amplitude is very time-consuming, resulting in a very limited dataset for performing accurate numerical integration. Second, the location of the bulk vertex might give rise to divergent integrands, as we will show in detail in this work, if there are small 4-simplices or close-to-null tetrahedra exist within the 1-5 move graph. Therefore, in order for the integration to be well defined, the measure of these divergent regions must be small enough to suppress the divergence. So far, this issue has not yet been fully investigated. 

Our immediate goal is first to understand the properties of the 1-5 Pachner move quantum geometry, specifically how the geometrical structure of the 1-5 Pachner move may affect the spinfoam amplitudes. To achieve this goal, we study extensively the impact of the position of the bulk vertex on the spinfoam amplitudes. We find that the two main factors are (1) the distribution of 4-simplex volumes in the 1-5 geometry, characterized by the standard deviation of individual 4-simplex volumes and the size of the smallest 4-simplex volume, and (2) how close certain tetrahedron normals are to the null direction. As we will also show in this paper, there is a clear distinction between the effects on spinfoam amplitudes generated by these two factors: the former factor impacts both the leading order and next-to-leading order contributions of the spinfoam amplitude, while the latter one mainly affects the next-to-leading order contribution, making it a quantum effect. The investigation in this paper will mainly be numerical, which has two main advantages. First, by employing numerical techniques, we can compute the large-$j$ asymptotic spinfoam amplitudes of 1-5 move of arbitrary samples and obtain actual value of the spinfoam (partial) amplitudes, which at the moment is only possible via numerical approach. Second, the dependence of the spinfoam action on the 4-dimensional geometry is generally very complicated analytically. Therefore, we will rely on numerical computation on a set of samples to give us straightforward insight to quickly and thoroughly pinpoint the main depending factors.

The structure of this paper is organized as follows: In section \ref{sec2}, we briefly review the general theory of defining the Lorentzian spinfoam EPRL amplitude in its general form, the saddle point approximation method, and the preparation for applying this method to study the large-$j$ asymptotics of spinfoam amplitudes. We also introduce the parametrization we used for the 1-5 move, along with the main algorithm of our computation. Section \ref{sec3} presents the main numerical results and corresponding analysis in detail, showing in several different setups how the topological location of the internal vertex can significantly impact the spinfoam amplitude with predictable patterns. Finally, in section \ref{sec4}, we summarize the main results obtained in this paper, discuss ongoing issues, and propose possible solutions. 

\section{Theoretical Background}\label{sec2}
\subsection{Spinfoam Amplitude and Poisson Summation}

A 4-dimensional simplicial complex $\mathcal{K}$ contains 4-simplices $v$, tetrahedra $e$, triangles $f$, line segments, and points. The internal and boundary triangles are denoted by $f=h$ and $f=b$. The LQG area spectrum indicates that the quantum area of triangle $f$ is given by $\mathfrak{a}_f = 8\pi\gamma G\hbar\sqrt{j_f (j_f + 1)}$ \cite{Rovelli:1994ge,Ashtekar:1996eg}, which can be approximated in the large-$j$ regime as $\mathfrak{a}_f \simeq \gamma j_f$, where we set $8\pi G\hbar = 1$. We also denote $j_h$, $j_b \in \mathbb{N}_0/2$ as the SU(2) spins assigned to internal and boundary triangles, respectively.

The Lorentzian EPRL spinfoam amplitude on the simplicial complex $\mathcal{K}$ has the following integral expression \cite{Conrady:2008ea, Conrady:2008mk1,Barrett:2009gg, Barrett:2009mw, Han:2011rf, Han:2011re1}:

\begin{equation}\label{action0}
    A(\mathcal{K})=\sum_{\left\{j_{h}\right\}}^{j^{\max }} \prod_{h} \boldsymbol{d}^{|V_h|+1}_{j_{h}} \int[\mathrm{d} g \mathrm{~d} \mathbf{z}] e^{S_{\rm SF}\left(j_{h}, g_{v e}, \mathbf{z}_{v f} ; j_{b}, \xi_{e b}\right)}, \quad[\mathrm{d} g \mathrm{~d} \mathbf{z}]=\prod_{(v, e)} \mathrm{d} g_{v e} \prod_{(v, f)} \mathrm{d} \Omega_{\mathbf{z}_{v f}},
\end{equation}
where $d_{j_h}=2j_h+1$, and the spinfoam action on the simplicial complex is expressed as:

\begin{equation}\label{action}
    \begin{aligned}
        S_{\rm SF} & =\sum_{e^{\prime}} j_{h} F_{\left(e^{\prime}, h\right)}+\sum_{(e, b)} j_{b} F_{(e, b)}^{\text {in/out }}+\sum_{\left(e^{\prime}, b\right)} j_{b} F_{\left(e^{\prime}, b\right)}^{\text {in out }}, \\
        F_{(e, b)}^{\text {out }} & =2 \ln \frac{\left\langle Z_{v e b}, \xi_{e b}\right\rangle}{\left\|Z_{v e b}\right\|}+i \gamma \ln \left\|Z_{v e b}\right\|^{2}, \\
        F_{(e, b)}^{\text {in }} & =2 \ln \frac{\left\langle\xi_{e b}, Z_{v^{\prime} e b}\right\rangle}{\left\|Z_{v^{\prime} e b}\right\|}-i \gamma \ln \left\|Z_{v^{\prime} e b}\right\|^{2}, \\
        F_{\left(e^{\prime}, f\right)} & =2 \ln \frac{\left\langle Z_{v e^{\prime} f}, Z_{v^{\prime} e^{\prime} f}\right\rangle}{\left\|Z_{v e^{\prime} f}\right\|\left\|Z_{v^{\prime} e^{\prime} f}\right\|}+i \gamma \ln \frac{\left\|Z_{v e^{\prime} f}\right\|^{2}}{\left\|Z_{v^{\prime} e^{\prime} f}\right\|^{2}}.
    \end{aligned}
\end{equation}
The spinfoam action $S$ is complex and linear to both the boundary and internal spins $j_b$ and $j_h$. Here, The boundary states are SU(2) coherent states $|j_b, \xi_{eb}\rangle$ where $\xi_{eb}= u_{eb}\triangleright(1, 0)^{\rm T}$, $u_{eb} \in \mathrm{SU(2)}$. $j_b$ and $\xi_{eb}$ are determined by the area and the 3-normal of the boundary triangle $b$, $V_h$ is the number of 4-simplices shared by the internal face $h$. $\mathrm{d}g_{ve}$ is the Haar measure on $\mathrm{SL(2,\mathbb{C})}$,

\be 
\mathrm{d} g=\frac{\mathrm{d} \alpha \mathrm{d} \alpha^{*} \mathrm{d} \beta \mathrm{d} \beta^{*} \mathrm{d} \gamma \mathrm{d} \gamma^{*} }{|\alpha|^{2}}, \quad \forall g=\left(\begin{array}{cc}
	\alpha & \beta \\
	\gamma & \delta
\end{array}\right)\in \Slc, 
\ee
and $\mathrm{d}\O_{\mathbf{z}_{v f}}$ is the scaling invariant measure on $\mathbb{CP}^1$:
\be 
\mathrm{d}\O_{\mathbf{z}_{v f}} 
&=&\frac{i}{2} \frac{\left(z_{0} \mathrm{~d} z_{1}-z_{1} \mathrm{~d} z_{0}\right) \wedge\left(\bar{z}_{0} \mathrm{~d} \bar{z}_{1}-\bar{z}_{1} \mathrm{~d} \bar{z}_{0}\right)}{\left\langle Z_{v e f}, Z_{v e f}\right\rangle\left\langle Z_{v e^{\prime} f}, Z_{v e^{\prime} f}\right\rangle}, \quad \forall \ {\bf z}_{vf}=(z_0,z_1)^{\rm T}, 
\ee where $Z_{v e f}=g_{v e}^{\rm T} \mathbf{z}_{v f}$ and $\mathbf{z}_{v f}$ is a 2-component spinor.  In (\ref{action}), $e$ and $e'$ are boundary and internal tetrahedra, respectively. In the dual complex $\mathcal{K}^*$, the orientation of $\partial f^*$ is outgoing from the vertex dual to $v$ and incoming to another vertex dual to $v'$, and the orientation of the face $f^*$ dual to $f$ induces $\partial f^*$’s orientation. As for the logarithms in the spinfoam action, we fix all the logarithms to be the principal values.

We would like to change the sum over $j_h$ in Eq. ({\ref{action0}}) to the integral, preparing for the stationary phase analysis. The idea is to apply a generalization of the Poisson summation formula \cite{BJBCrowley_1979}
\[
\sum_{n=0}^{n_{\rm max}} f(n)=\sum_{k \in \mathbb{Z}} \int_{-\epsilon}^{n_{\rm max}+1-\epsilon} \mathrm{d} n f(n) \,{e}^{2\pi i k n},
\]
where $\epsilon\in(0,1)$. Identifying $n=2j$, $n_{\rm max}=2j^{\rm max}$ and $f(n)$ to be the summand in the spinfoam amplitude, we obtain the following expression of $A(\ck)$
\be
A(\ck)&=&\sum_{\{k_h\in\mathbb{Z}\}}\int\limits_{-\epsilon}^{2j^{\rm max}+1-\epsilon} 
\prod_f\mathrm{d}(2 j_{f}) \,\prod_h\mu_h(j_f)\int [\mathrm{d} g \mathrm{~d} \mathbf{z}]\, e^{ S_{\rm SF}^{(k)}}, \label{integralFormAmp1}\\
&&S_{\rm SF}^{(k)}=S_{\rm SF}+4\pi i \sum_f j_f k_f
\ee

To probe the large-$j$ regime, we scale boundary spins $j_b\to \l j_b$ with any $\l\gg1$, and make the change of variables $j_h\to\l j_h$. We also scale $j^{\rm max}$ by $j^{\rm max}\to \l j^{\rm max}$. Then, $A(\ck)$ is given by  
\be
A(\ck)&=&\sum_{\{k_h\in\mathbb{Z}\}}\int\limits_{-\epsilon/\lambda }^{2j^{\rm max}+({1-\epsilon})/{\l}} 
\prod_f\mathrm{d}(2 j_{f}) \,\prod_h\mu_h(\l j_f)\int [\mathrm{d} g \mathrm{~d} \mathbf{z}]\, e^{\l S_{\rm SF}^{(k)}}\label{integralFormAmp22}
\ee
which is used in our discussion. Here, we focus on the amplitude with all $k_h=0$ since it often corresponds to the dominant contribution given suitable boundary condition, as suggested in the numerical results in, e.g., \cite{Han:2021kll}.

The asymptotical behavior of $A(\mathcal{K})$ in the large-$j$ regime can be studied by performing generalized stationary phase approximation analysis guided by Hörmander’s theorem 7.7.5 \cite{hormander2015analysis}:
\begin{theorem}\label{theorem111}
Let K be a compact subset in $\mathbb{R}^n$, X an open neighborhood of K, and k a positive integer. If: (1) the complex functions $u \in C_0^{2 k}(K)$, $f \in C^{3 k+1}(X)$ and $\operatorname{Im} f(x) \geq 0 \quad\forall x\in X$, (2) there is a unique point $x_0 \in K$ satisfying $\operatorname{Im}\left(S\left(x_0\right)\right)=0$, $f^{\prime}\left(x_0\right)=0$, $\operatorname{det}\left(f^{\prime \prime}\left(x_0\right)\right) \neq 0$ (where $f''$ denotes the Hessian matrix), and $f^{\prime}(x) \neq 0 \quad\forall x\in K \backslash\left\{x_0\right\}$, then we have the following estimation:
\begin{equation}\label{saddle1}
\left|\int_K u(x) e^{i \lambda f(x)} d x-e^{i \lambda f \left(x_0\right)}\left[\operatorname{det}\left(\frac{\lambda f^{\prime \prime}\left(x_0\right)}{2 \pi i}\right)\right]^{-\frac{1}{2}} \sum_{s=0}^{k-1}\left(\frac{1}{\lambda}\right)^s L_s u\left(x_0\right)\right| \leq C\left(\frac{1}{\lambda}\right)^k \sum_{|\alpha| \leq 2 k} \sup \left|D^\alpha u\right| .
\end{equation}
Here, the constant $C$ is bounded when $f$ stays in a bounded set in $C^{3 k+1}(X)$. We use the standard multi-index notation $\alpha=\left\langle\alpha_1, \ldots, \alpha_n\right\rangle$ and:
\begin{equation}
D^\alpha=(-i)^\alpha \frac{\partial^{|\alpha|}}{\partial x_1^{\alpha_1} \ldots \partial x_n^{\alpha_n}}, \quad \text { where } \quad|\alpha|=\sum_{i=1}^n \alpha_i,
\end{equation}
 $L_s u\left(x_0\right)$ is defined as:
\begin{equation}
L_s u\left(x_0\right)=i^{-s} \sum_{l-m=s} \sum_{2 l \geq 3 m} \frac{(-1)^l 2^{-l}}{l ! m !}\left[\sum_{a, b=1}^n H_{a b}^{-1}\left(x_0\right) \frac{\partial^2}{\partial x_a \partial x_b}\right]^l\left(g_{x_0}^m u\right)\left(x_0\right),
\end{equation}
where $H(x)=f^{\prime \prime}(x)$ denotes the Hessian matrix and the function $g_{x_0}(x)$ is given by:
\begin{equation}
g_{x_0}(x)=f(x)-f\left(x_0\right)-\frac{1}{2} H^{a b}\left(x_0\right)\left(x-x_0\right)_a\left(x-x_0\right)_b,
\end{equation}
satisfying $g_{x_0}\left(x_0\right)=g_{x_0}^{\prime}\left(x_0\right)=g_{x_0}^{\prime \prime}\left(x_0\right)=0$. For each $s$, $L_s$ is a differential operator of order $2s$ acting on $u(x)$.
\end{theorem}

According to this Theorem, the spinfoam amplitude in (\ref{integralFormAmp22}) can be computed as an $1/\lambda$ asymptotic series at the real critical points $\mathring{x}=\{\mathring{j}_{h}, \mathring{g}_{v e}, \mathring{\mathbf{z}}_{v f}\}$, which are the solution of the critical equations,
\begin{equation}
    \begin{split}
        \mathrm{Re}(S)&=\partial_{g_{ve}}S=\partial_{\textbf{z}_{vf}}S=0,\\
        \partial_{j_h}S&=4\pi ik_h, k_h\in\mathbb{Z},
    \end{split}
\end{equation}
where $\partial_{j_h}S=4\pi ik_h$ with $k_h \neq 0$ coming from higher-order expansions of the Poisson summation formula. In this paper, we focus only on the $\partial_{j_h}S=0$ case. Results from \cite{Han:2013gna,Han:2020fil} show that the existence of the real critical point in \eqref{integralFormAmp22} depends on the boundary condition. As one of the results, the real critical point satisfying a nondegeneracy condition endows a flat Regge geometry on $\ck$ with certain 4-simplex orientations. $S$ evaluated at critical points gives the Regge action on the simplicial complex. In the next section, we review the simplicial complex of the 1-5 move and prepare for the numerical computation of the spinfoam amplitude on the 1-5 move employing this theorem.

\subsection{Real parametrization of 1-5 Pachner Move}
\begin{figure}[H]
\centering
\begin{subfigure}[b]{0.4\textwidth}
\centering
\includegraphics[width=\textwidth]{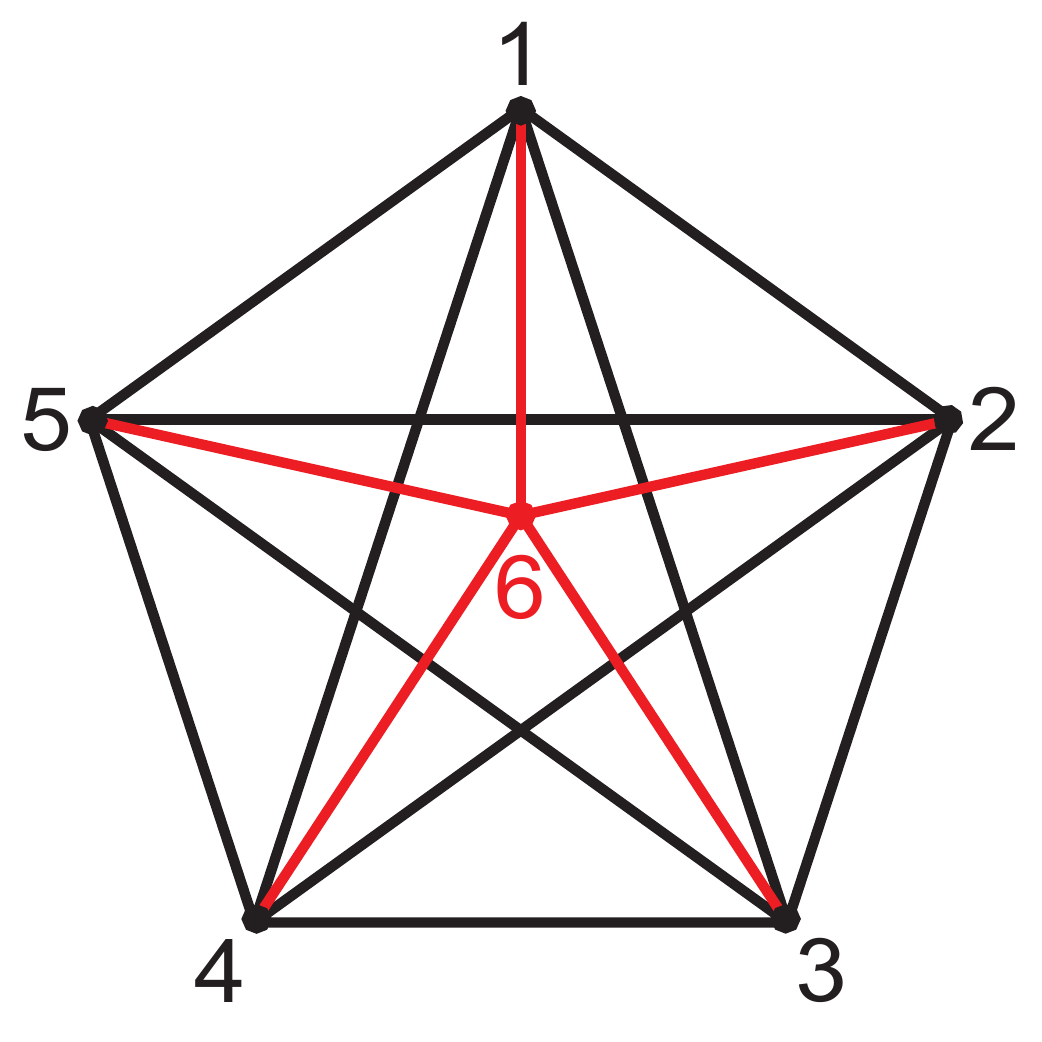}
\caption{}
\end{subfigure}
\begin{subfigure}[b]{0.4\textwidth}
\centering
\includegraphics[width=\textwidth]{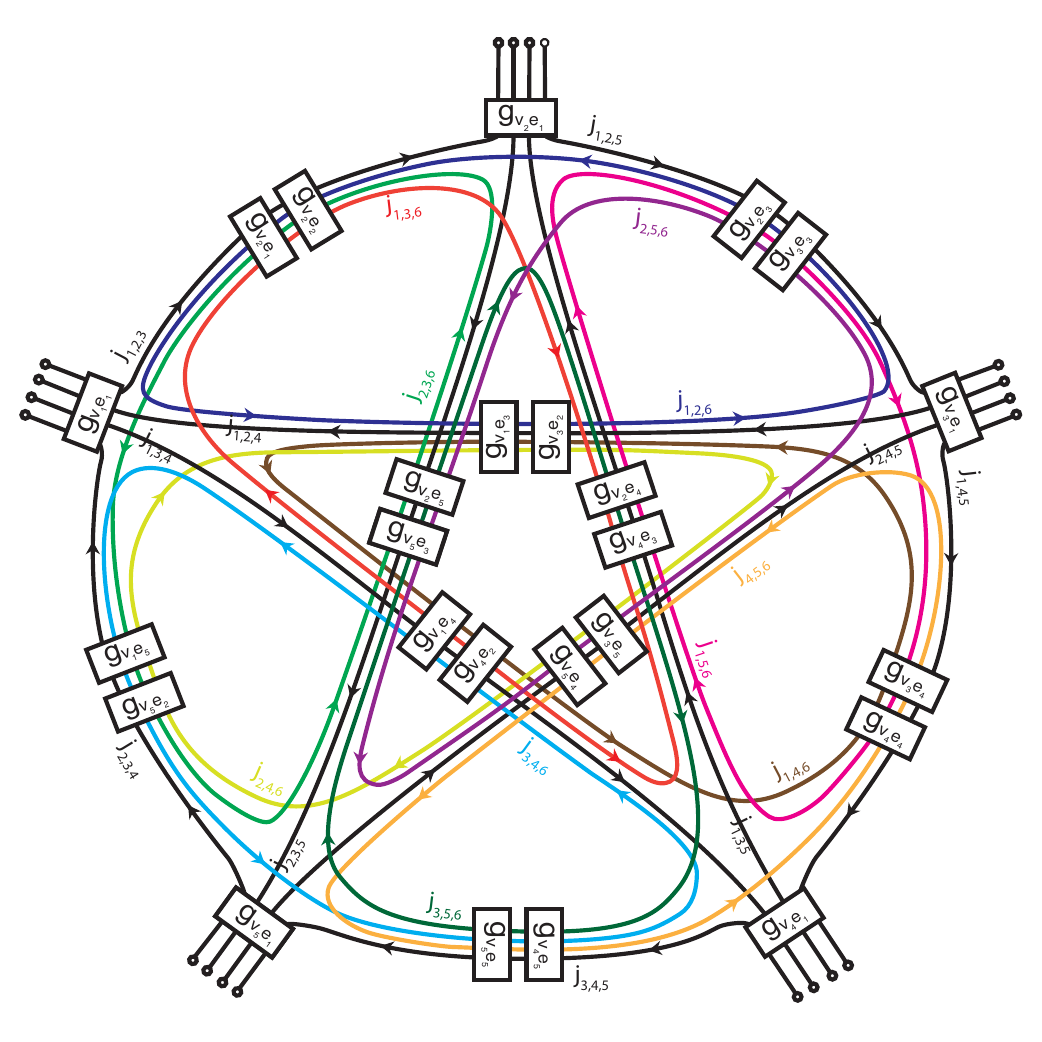}
\caption{}
\end{subfigure}
\caption{(a) Illustration of $\sigma_{1-5}$: the boundary 4-simplex $S_{12345}$ is divided into five 4-simplices $S_{12346}$, $S_{12356}$, $S_{12456}$, $S_{13456}$ and $S_{23456}$. $\sig_{\text{1-5}}$ has 10 boundary triangles, 10 internal triangles and 5 internal segments $l_{I6}$ with $I=1,\cdots,5$ (red). (b) The dual cable diagram of the $\sigma_{1-5}$ spinfoam amplitude: The boxes correspond to tetrahedra carrying $g_{ve} \in \mathrm{SL(2, \mathbb{C})}$. The strands stand for triangles carrying spins $j_f$. The strand with the same color belonging to a different dual vertex corresponds to the triangle shared by the different 4-simplices. The circles as the endpoints of the strands carry boundary states $|j_b, \xi_{eb}\rangle$. The arrows represent orientations. This figure is adapted from \cite{Banburski:2014cwa}.}\label{1-5move1}
\end{figure}

$\sigma_{1-5}$ is the simplicial complex of the 1-5 Pachner move, refining a 4-simplex into five 4-simplices. The initial configuration of 1-5 move is a 4-simplex labeled by five points, and the final configuration contains five 4-simplices by adding a point 6 inside the initial boundary 4-simplex and connecting point 6 to other 5 points of the 4-simplex by five line segments $l_{16},l_{26}, \cdots,l_{56}$, see FIG. \ref{1-5move1} (a). The dual cable diagram of $\sigma_{1-5}$ amplitude is in FIG. \ref{1-5move1} (b). $\sigma_{1-5}$ consists of five internal segments, 10 boundary triangles $b$ (dual to black strands in FIG. \ref{1-5move1} (b)) and 10 internal triangles $h$ \footnote{$j_f$ appeared in FIG. \ref{1-5move1} (b) can be categorized as: \\
$j_b=\{j_{1,2,3},j_{1,2,4},j_{1,2,5},j_{1,3,4},j_{1,3,5},j_{1,4,5},j_{2,3,4},j_{2,3,5},j_{2,4,5},j_{3,4,5}\}$\\
$j_h=\{j_{1,2,6},j_{1,3,6},j_{1,4,6},j_{1,5,6},j_{2,3,6},j_{2,4,6},j_{2,5,6},j_{3,4,6},j_{3,5,6},j_{4,5,6}\}$\\} (dual to colored loops in FIG. \ref{1-5move1} (b)). 

The coordinates of the points on $\sigma_{1-5}$ are: 
\be 
\begin{gathered}
P_{1}=(0, 0, 0, 0),\quad  P_{2}=(0, -\frac{2\sqrt{10}}{3^{\frac{3}{4}}}, -\frac{\sqrt{5}}{3^{\frac{3}{4}}}, -\frac{\sqrt{5}}{3^{\frac{1}{4}}}), \quad P_{3}=(0, 0, 0, -\frac{2\sqrt{5}}{3^{\frac{1}{4}}}), \\
P_{4}=(-\frac{1}{3^{\frac{1}{4}}\sqrt{10}}, -\frac{\sqrt{10}}{2\times 3^{\frac{3}{4}}}, -\frac{\sqrt{5}}{3^{\frac{3}{4}}},  -\frac{\sqrt{5}}{3^{\frac{1}{4}}}),\quad P_{5}=(0, 0, -3^{\frac{1}{4}}\sqrt{5}, -\frac{\sqrt{5}}{3^{\frac{1}{4}}}), \\
P_{6}=\sum_{i=1}^5 a_i\,P_i,\quad 0\leq a_i\leq 1. \label{coordinates}
\end{gathered}
\ee
$P_{1}, \cdots, P_{6}$ determines a flat Regge geometry on $\sigma_{1-5}$. Five Lorentzian 4-simplices $v_1=S_{12346}$, $v_2=S_{12356}$, $v_3=S_{12456}$, $v_4=S_{13456}$, $v_5=S_{23456}$ can then be obtained, and the corresponding tetrahedra and triangles are all spacelike. With the coordinates in (\ref{coordinates}), we use the algorithm in \cite{Han:2024lti} to compute the boundary data $(j_b,\xi_{eb})$ and the real critical points $\mathring{x}=(\mathring{j}_h,\mathring{g}_{ve},\mathring{z}_{vf})$.

The spinfoam action of $\sigma_{1-5}$ has the following continuous gauge freedom. We need to fix these gauge freedoms and perform the real parametrization to prepare for numerical computation. The gauge freedoms in the spinfoam action and the corresponding gauge fixing are as follows:
\begin{itemize}
    \item At each $v$, there is the $\mathrm{SL(2,\mathbb{C})}$ gauge freedom $g_{v e} \mapsto x_v^{-1} g_{v e}$, $\mathbf{z}_{v f} \mapsto x_v^{\dagger} \mathbf{z}_{v f}$, $x_v \in \mathrm{SL}(2, \mathbb{C})$. We fix $g_{ve}$ to be a constant $\mathrm{SL(2,\mathbb{C})}$ matrix for $(v,e)\in\{(v_1,e_1),(v_2,e_1),(v_3,e_1),(v_4,e_1),(v_5,e_1)\}$ \footnote{For the same tetrahedron, the correspondence between label $(v,e)$ and $(P_{i},P_{j},P_{k},P_{l})$ is:\\ $\{(v_1,e_1),(v_1,e_2),(v_1,e_3),(v_1,e_4),(v_1,e_5)\} = \{(P_1,P_2,P_3,P_4),(P_1,P_2,P_3,P_6),(P_1,P_2,P_4,P_6),(P_1,P_3,P_4,P_6),(P_2,P_3,P_4,P_6)\}$,\\ $\{(v_2,e_1),(v_2,e_2),(v_2,e_3),(v_2,e_4),(v_2,e_5)\} = \{(P_1,P_2,P_3,P_5),(P_1,P_2,P_3,P_6),(P_1,P_2,P_5,P_6),(P_1,P_3,P_5,P_6),(P_2,P_3,P_5,P_6)\}$,\\ $\{(v_3,e_1),(v_3,e_2),(v_3,e_3),(v_3,e_4),(v_3,e_5)\} = \{(P_1,P_2,P_4,P_5),(P_1,P_2,P_4,P_6),(P_1,P_2,P_5,P_6),(P_1,P_4,P_5,P_6),(P_2,P_4,P_5,P_6)\}$,\\ $\{(v_4,e_1),(v_4,e_2),(v_4,e_3),(v_4,e_4),(v_4,e_5)\} = \{(P_1,P_3,P_4,P_5),(P_1,P_3,P_4,P_6),(P_1,P_3,P_5,P_6),(P_1,P_4,P_5,P_6),(P_3,P_4,P_5,P_6)\}$,\\ $\{(v_5,e_1),(v_5,e_2),(v_5,e_3),(v_5,e_4),(v_5,e_5)\} = \{(P_2,P_3,P_4,P_5),(P_2,P_3,P_4,P_6),(P_2,P_3,P_5,P_6),(P_2,P_4,P_5,P_6),(P_3,P_4,P_5,P_6)\}$. }. The amplitude is independent of the choices of constant matrices.
    \item At each $e$, there is the $\mathrm{SU}(2)$ gauge freedom: $g_{v^{\prime} e} \mapsto g_{v^{\prime} e} h_e^{-1}$, $g_{v e} \mapsto g_{v e} h_e^{-1}$, $h_e \in \mathrm{SU}(2)$. To remove the gauge freedom, we set one of the group elements $g_{v'e}$ along the edge $e$ to be the lower triangular matrix. Indeed, any $g \in \operatorname{SL}(2, \mathbb{C})$ can be decomposed as $g=k h$ with $h \in \mathrm{SU}(2)$ and $k \in K$, where $K$ is the subgroup of lower triangular matrices:
    \begin{equation}
    K=\left\{k=\left(\begin{array}{cc}
    \lambda^{-1} & 0 \\
    \mu & \lambda
    \end{array}\right), \lambda \in \mathbb{R} \backslash\{0\}, \mu \in \mathbb{C}\right\}
    \end{equation}
    We use the gauge freedom to set $g_{v^{\prime} e} \in K$. On $\sigma_{1-5}$, we fix the group element $g_{ve}$ for the bulk tetrahedra $(v,e)\in\{(v_1,e_2),(v_1,e_3),(v_1,e_4),(v_1,e_5),(v_2,e_3),(v_2,e_4),(v_2,e_5),(v_3,e_4),(v_3,e_5),(v_4,e_5)\}$ to be the lower triangular matrix.
    \item $\mathbf{z}_{v f}$ can be computed by $g_{v e}$ and $\xi_{e f}$ up to a complex scaling: $\mathbf{z}_{v f} \propto_{\mathbb{C}}\left(g_{v e}^{T}\right)^{-1} \xi_{e f}$. Each $\mathbf{z}_{v f}$ has the scaling gauge freedom $\mathbf{z}_{v f} \mapsto \lambda_{v f} \mathbf{z}_{v f}, \lambda_{v f} \in \mathbb{C}$. We fix the gauge by setting the first component of $\mathbf{z}_{v f}$ to 1. Then, the real critical point $\stackrel{\circ}{\mathbf{z}}_{v f}$ is in the form of $\stackrel{\circ}{\mathbf{z}}_{v f}=\left(1, \stackrel{\circ}{\alpha}_{v f}\right)^{T}$, where $\stackrel{\circ}{\alpha}_{v f} \in \mathbb{C}$.
\end{itemize}

Now, we introduce the real parametrizations of all the integration variables.  For the variables related with $\mathrm{SL}(2,\mathbb{C})$ group elements we have:
\begin{itemize}
  \item As $(v,e)\in\{(v_1,e_1),(v_2,e_1),(v_3,e_1),(v_4,e_1),(v_5,e_1)\}, g_{ve}=g_{ve}^{0}$.
  \item As $(v,e)\in\{(v_1,e_2),(v_1,e_3),(v_1,e_4),(v_1,e_5),(v_2,e_3),(v_2,e_4),(v_2,e_5),(v_3,e_4),(v_3,e_5),(v_4,e_5)\}$, $g_{ve}$ is gauge-fixed to be an lower triangular matrix:
\end{itemize}
\begin{equation}
    g_{ve}=\left(\begin{array}{cc}
    1+x_{ve}^{1} & 0 \\
    x_{ve}^{2}+i y_{ve}^{2} & \mu_{ve}
    \end{array}\right),
\end{equation}
here, $\mu_{ve}$ is determined by $\operatorname{det} g_{ve}=1$.
\begin{itemize}
  \item As $(v,e)\in\{(v_2,e_2),(v_3,e_2),(v_3,e_3),(v_4,e_2),(v_4,e_3),(v_4,e_4),(v_5,e_2),(v_5,e_3),(v_5,e_4),(v_5,e_5)\}$, $g_{ve}$ is parameterized by
\end{itemize}
\begin{equation}
    g_{ve}=\left(\begin{array}{cc}
    1+x_{ve}^{1}+i y_{ve}^{1} & x_{ve}^{2}+i y_{ve}^{2} \\
    x_{ve}^{3}+i y_{ve}^{3} & \mu_{ve}
    \end{array}\right)
\end{equation}
Each spinor is parametrized by two real parameters:
\begin{equation}
    \mathbf{z}_{vf}=\left(1, x_{vf}+i y_{vf}\right) .
\end{equation}
For the internal spin, we parametrize it by one real parameter $\frak{j}_h$.

With these parametrizations, the measures $\mathrm{~d}g_{ve}$ and $\mathrm{~d}\Omega_{z_{vf}}$ can be calculated as: \cite{ruhl1970lorentz,Conrady:2008ea,Han:2020fil}
\begin{equation}
    \begin{split}
        &\mathrm{~d}g_{ve}=\frac{1}{(8\pi^3}(1+x^1_{ve})\mathrm{~d}x^1_{ve}\mathrm{~d}x^2_{ve}\mathrm{~d}y^2_{ve}, \\
        &\quad\qquad\forall (v,e)\in\{(v_1,e_2),(v_1,e_3),(v_1,e_4),(v_1,e_5),(v_2,e_3),(v_2,e_4),(v_2,e_5),(v_3,e_4),(v_3,e_5),(v_4,e_5)\}\\
        &\mathrm{~d}g_{ve}=\frac{1}{16\pi^4}\frac{\mathrm{~d}x^1_{ve}\mathrm{~d}x^2_{ve}\mathrm{~d}x^3_{ve}\mathrm{~d}y^1_{ve}\mathrm{~d}y^2_{ve}\mathrm{~d}y^3_{ve}}{|1+x^1_{ve}+i y^1_{ve}|^2},\\
        &\quad\qquad\forall (v,e)\in\{(v_2,e_2),(v_3,e_2),(v_3,e_3),(v_4,e_2),(v_4,e_3),(v_4,e_4),(v_5,e_2),(v_5,e_3),(v_5,e_4),(v_5,e_5)\}\\
        &\mathrm{~d}\Omega_{\textbf{z}_{vf}}=\frac{\mathrm{~d}x_{vf}\mathrm{~d}y_{vf}}{\langle Z_{vef},Z_{vef}\rangle\langle Z_{ve'f},Z_{ve'f}\rangle},
    \end{split}
\end{equation}
where $e$ and $e'$ are the two tetrahedron sharing $f$ inside $v$.

Unlike a single 4-simplex, for which, up to this step, we can move on to compute the full next-to-leading order spinfoam amplitude, the flat geometry on $\sigma_{1-5}$ is not unique. Given a fixed boundary 4-simplex, the position of $P_{6}$ can move continuously in $\mathbb{R}^{4}$, which leads to the continuous family of flat geometries on $\sigma_{1-5}$. This continuous family of flat geometries results in a continuous family of real critical points, which implies that all these real critical points lead to degenerate Hessian matrices. To tackle this problem, we take the following additional procedure introduced in \cite{Han:2021kll}: We select four internal spins $j_{156}, j_{256}, j_{356}$ and $j_{456}$ independently as the outermost integration variables, where the subindex $abc$ in $j_{abc}$ stands for the three endpoints of the triangle whose area is labeled by $j_{abc}$. The full integral is then denoted by:
\begin{equation}\label{int1}
    \begin{aligned}
    A\left(\sigma_{1-5}\right) & =\int_{\mathbb{R}^{4}} \mathrm{~d} j_{156} \mathrm{~d} j_{256} \mathrm{~d} j_{356} \mathrm{~d} j_{456}\cdot \mathcal{Z}^J_{\sigma_{1-5}}\left(j_{156}, j_{256}, j_{356}, j_{456}\right), \\
    &=\int_{\mathbb{R}^{4}} \mathrm{~d} x_{v_6} \mathrm{~d} y_{v_6} \mathrm{~d} z_{v_6} \mathrm{~d} t_{v_6} \mathcal{Z}_{\sigma_{1-5}}\left(x_{v_6},y_{v_6},z_{v_6},t_{v_6}\right), \\
    \mathcal{Z}_{\sigma_{1-5}} & =\int_{\mathbb{R}^{6}} \prod_{\bar{h}=1}^{6} \mathrm{~d} j_{\bar{h}} \prod_{h=1}^{10} 2 \lambda \tau_{\left[-\epsilon, \lambda j^{\max }+\epsilon\right]}\left(\lambda j_{h}\right) \cdot J_{jx}\cdot \int[\mathrm{d} g \mathrm{~d} \mathbf{z}] e^{\lambda S},
    \end{aligned}
\end{equation}
where $x_{v_6}$, $y_{v_6}$, $z_{v_6}$, $t_{v_6}$ are the coordinates of the internal vertex, $J_{jx}$ is the Jacobian determinant between $j_{156}$, $j_{256}$, $j_{356}$, $j_{456}$ and $x_{v_6}$, $y_{v_6}$, $z_{v_6}$, $t_{v_6}$. This change of variable is possible due to the correspondence between internal face areas and these four $j$ variables. The other six internal spins $j_{126}$, $j_{136}$, $j_{146}$, $j_{236}$, $j_{246}$, $j_{346}$ are denoted by $j_{\bar{h}} (\bar{h}=1,2,\cdots,6)$. The partial amplitude $\mathcal{Z}_{\sigma_{1-5}}$ contains integrals of the following 196 real variables:
\begin{equation}
    \vec{x} = (x_{ve}^1,x_{ve}^2,x_{ve}^3, y_{ve}^1,y_{ve}^2,y_{ve}^3, x_{vf},y_{vf}, \frak{j}_h)
\end{equation}

For an arbitrary combination of $j_{156}, j_{256}, j_{356}, j_{456}$ that preserves the same $\sigma_{1-5}$ geometrical structure such that all tetrahedra are spacelike and $P_6$ remains to be the internal point, a unique real critical point $\{\mathring{j}_{h}, \mathring{g}_{ve}, \mathring{z}_{vef}\}$ can be determined \cite{Han:2024lti}, up to each 4-simplex orientation defined in \cite{Han:2011re1}. In this paper, we choose the orientation $(+,+,+,+,+)$ for all the samples we study. As a result, the large-$j$ expansions of the partial amplitude $\mathcal{Z}_{\sigma_{1-5}}$ can be computed using Theorem \ref{theorem111}.

\subsection{Improved Numerical Algorithm}
Following the convention in Theorem \ref{theorem111}, we rewrite the partial amplitude in the form of: 
\begin{equation}
\mathcal{Z}_{\sigma_{1-5}} = \int \prod_{i=1}^{196}dx_i u\left(\vec{\mathbf{x}}\right)  e^{\lambda S\left(\vec{\mathbf{x}}\right)}=\int \prod_{i=1}^{196}dx_i u\left(\vec{\mathbf{x}}\right)  e^{i\lambda \tilde{S}\left(\vec{\mathbf{x}}\right)}.
\end{equation}
Here, $\tilde{S}\left(\vec{\mathbf{x}}\right) := -iS\left(\vec{\mathbf{x}}\right)$, Hessian matrix $H_{ij}(\vec{\mathbf{x}})=\partial_i\partial_j\tilde{S}(\vec{\mathbf{x}})$. The leading and the next-to-leading order terms in Eq.(\ref{saddle1}) correspond to $s=0$ and $s=1$. We denote the critical point as $x_0:=\{\mathring{\frak{j}}_{h}, \mathring{g}_{ve}, \mathring{z}_{vef}\}$. In (\ref{saddle1}), the expression of $L_s u\left(x_0\right)$ sums a finite number of terms for each $s$.

At $s=0$, the corresponding term for $L_{s=0} u(\overrightarrow{\mathbf{x}})$ is:
\begin{equation}
I_0=u(x_0).
\end{equation}

At $s=1$, our numerical computation to obtain the next-to-leading order quantum corrections is currently done in Mathematica, employing in principle the same numerical method as described in \cite{Han:2020fil} with some additional optimizations, which is described below: the possible combinations of $(m, l)$ satisfying $2 l>3 m$ are $(0,1)$, $(1,2)$, $(2,3)$, which can be computed separately as:
\begin{itemize}
    \item $(m, l)=(0,1)$:
    \begin{equation}
I_1=-\frac{1}{2 i}\left[\sum_{i, j=1}^{196} H_{i j}^{-1}(x_0) \frac{\partial^2 u(x_0)}{\partial x_i \partial x_j}\right],
\end{equation}
where $x_i$ spans over each of the 196 real variables in our numerical model. This term is easy to compute, and the symmetric properties of both the Hessian matrix and second-order derivatives are used to optimize the computation.
\item $(m, l)=(1,2)$:
\begin{equation}
\begin{aligned}
I_2= & \frac{1}{8 i}\left[\sum_{i, j=1}^{196} H_{i j}^{-1}(x_0) \frac{\partial^2}{\partial x_i \partial x_j}\right]\left[\sum_{k, l=1}^{196} H_{k l}^{-1}(x_0) \frac{\partial^2}{\partial x_k \partial x_l}\right]\left(g_{x_0} u\right)(x_0) \\
= & \frac{1}{8 i}\left[\sum_{i, j, k, l=1}^{196} H_{i j}^{-1} H_{k l}^{-1} \frac{\partial^4}{\partial x_i \partial x_j \partial x_k \partial x_l}\right]\left(g_{x_0} u\right)(x_0) \\
= & \frac{1}{8 i} \sum_{i, j, k, l=1}^{196} H_{i j}^{-1} H_{k l}^{-1}\left[\frac{\partial^3 g_{x_0}(x_0)}{\partial x_i \partial x_j \partial x_k} \frac{\partial u(x_0)}{\partial x_l}+\frac{\partial^3 g_{x_0}(x_0)}{\partial x_i \partial x_j \partial x_l} \frac{\partial u(x_0)}{\partial x_k}+\frac{\partial^3 g_{x_0}(x_0)}{\partial x_j \partial x_k \partial x_l} \frac{\partial u(x_0)}{\partial x_i}\right. \\
& \left.\quad+\frac{\partial^3 g_{x_0}(x_0)}{\partial x_i \partial x_k \partial x_l} \frac{\partial u(x_0)}{\partial x_j}+\frac{\partial^4 g_{x_0}(x_0)}{\partial x_i \partial x_j \partial x_k \partial x_l} u(x_0)\right] .
\end{aligned}
\end{equation}
where we define:
\begin{equation}
g_{x_0}(\overrightarrow{\mathbf{x}}):=\tilde{S}(\overrightarrow{\mathbf{x}})-\tilde{S}(x_0)-\frac{1}{2} \sum_{i, j=1}^{196} H_{i j}(x_0) x_i x_j.
\end{equation}
\item $(m, l)=(2,3)$:
\begin{equation}
\begin{aligned}
I_3= & -\frac{1}{96 i}\left[\sum_{i, j, k, l, m, n=1}^{196} H_{i j}^{-1} H_{k l}^{-1} H_{m n}^{-1} \frac{\partial^6}{\partial x_i \partial x_j \partial x_k \partial x_l \partial x_m \partial x_n}\right]\left(g_{x_0}^2 u\right) \\
& =-\frac{1}{48 i} \sum_{i, j, k, l, m, n=1}^{196} H_{i j}^{-1} H_{k l}^{-1} H_{m n}^{-1}\left[\frac{\partial^3 g_{x_0}(x_0)}{\partial x_j \partial x_k \partial x_l } \frac{\partial^3 g_{x_0}(x_0)}{\partial x_i \partial x_m \partial x_n}\right. \\
& +\frac{\partial^3 g_{x_0}(x_0)}{\partial x_j \partial x_k \partial x_m} \frac{\partial^3 g_{x_0}(x_0)}{\partial x_i \partial x_l \partial x_n}+\frac{\partial^3 g_{x_0}(x_0)}{\partial x_j \partial x_k \partial x_n} \frac{\partial^3 g_{x_0}(x_0)}{\partial x_i \partial x_l \partial x_m} +\frac{\partial^3 g_{x_0}(x_0)}{\partial x_j \partial x_l \partial x_m} \frac{\partial^3 g_{x_0}(x_0)}{\partial x_i \partial x_k \partial x_n}+\frac{\partial^3 g_{x_0}(x_0)}{\partial x_j \partial x_l \partial x_n} \frac{\partial^3 g_{x_0}(x_0)}{\partial x_i \partial x_k \partial x_m} \\
& +\frac{\partial^3 g_{x_0}(x_0)}{\partial x_j \partial x_m \partial x_n} \frac{\partial^3 g_{x_0}(x_0)}{\partial x_i \partial x_k \partial x_l}+\frac{\partial^3 g_{x_0}(x_0)}{\partial x_k \partial x_l \partial x_m} \frac{\partial^3 g_{x_0}(x_0)}{\partial x_i \partial x_j \partial x_n}+\frac{\partial^3 g_{x_0}(x_0)}{\partial x_k \partial x_l \partial x_n} \frac{\partial^3 g_{x_0}(x_0)}{\partial x_i \partial x_j \partial x_m}+\frac{\partial^3 g_{x_0}(x_0)}{\partial x_k \partial x_m \partial x_n} \frac{\partial^3 g_{x_0}(x_0)}{\partial x_i \partial x_j \partial x_l} \\
& \left.+\frac{\partial^3 g_{x_0}(x_0)}{\partial x_l \partial x_m \partial x_n} \frac{\partial^3 g_{x_0}(x_0)}{\partial x_i \partial x_j \partial x_k}\right] u\left(x_0\right) .
\end{aligned}
\end{equation}
\end{itemize}
For computing $I_2$ and $I_3$, note first that the {most of the terms in $\frac{\partial^3 g_{x_0}(x_0)}{\partial x_i \partial x_j \partial x_k}$ and $\frac{\partial^4 g_{x_0}(x_0)}{\partial x_i \partial x_j \partial x_k \partial x_l}$ are zero. Thus we only need to compute the terms in which $\frac{\partial^3 g_{x_0}(x_0)}{\partial x_i \partial x_j \partial x_k}$ and $\frac{\partial^4 g_{x_0}(x_0)}{\partial x_i \partial x_j \partial x_k \partial x_l}$ are non-zero. In practice, this process begins with an index table that keeps track of indices of all non-zero derivatives. This index table is stored uniquely in the order $(i,j,k,\dots),i\leq j\leq k\leq \dots\leq 196$ such that the symmetric properties of the derivatives are fully considered. For each element of the index table, all possible permutations of the corresponding Hessian matrices are considered. To further save performance, we fully investigate algorithm consistency in terms of parallelization and successfully generalized the index table to different types of permutations firsthand. For example, all possible set of indices of $\frac{\partial^4 g_{x_0}(x_0)}{\partial x_i \partial x_j \partial x_k \partial x_l}$ fall into the following categories: $(1,2,3,4)$, $(1,1,2,3)$, $(1,2,2,3)$, $(1,2,3,3)$, $(1,1,1,2)$, $(1,2,2,2)$, $(1,1,1,1)$. For each of these cases, the number of repetitions due to the symmetry of the Hessian matrix during contraction is computed and stored first as a global table. Then, for each actual set of derivative indices, we take the values of the corresponding Hessian matrices by identifying the index pattern with one of the index types listed above. This significantly reduces computational cost, enabling much faster computation.

Also, it is extremely time and memory consuming to directly compute every possible $\frac{\partial^4 g_{x_0}(x_0)}{\partial x_i \partial x_j \partial x_k \partial x_l}$ due to the large number of variables in our case. The computation is greatly accelerated by first determining all analytically non-zero third-order derivatives and computing only the derivative of these non-zero third-order derivatives to obtain non-zero $\frac{\partial^4 g_{x_0}(x_0)}{\partial x_i \partial x_j \partial x_k \partial x_l}$.

Finally, the partial spinfoam amplitude at the real critical point can be computed up to its next-to-leading order correction as:
\begin{equation}
\begin{aligned}
\mathcal{Z}_{\sigma_{1-5}} & =\mathcal{Z}_{\sigma_{1-5}}^{ 0}+\mathcal{Z}_{\sigma_{1-5}}^{ 1}+O\left(\frac{1}{\lambda^2}\right), \\
\mathcal{Z}_{\sigma_{1-5}}^{ 0} & =e^{i \lambda \tilde{S}\left(\overrightarrow{\mathbf{x}}_0\right)}\left[\operatorname{det}\left(\frac{\lambda S^{\prime \prime}\left(\overrightarrow{\mathbf{x}}_0\right)}{2 \pi i}\right)\right]^{-\frac{1}{2}} u\left(\overrightarrow{\mathbf{x}}_0\right), \\
\mathcal{Z}_{\sigma_{1-5}}^{ 1} & =e^{i \lambda \tilde{S}\left(\overrightarrow{\mathbf{x}}_0\right)}\left[\operatorname{det}\left(\frac{\lambda S^{\prime \prime}\left(\overrightarrow{\mathbf{x}}_0\right)}{2 \pi i}\right)\right]^{-\frac{1}{2}} \frac{1}{\lambda}\left(I_1+I_2+I_3\right)\left(\overrightarrow{\mathbf{x}}_0\right) .
\end{aligned}
\end{equation}

\section{Main Results}\label{sec3}
\subsection{Monte Carlo Sampling}
To numerically compute the total integral in eqn. (\ref{int1}), the integrand $\mathcal{Z}_{\sigma_{1-5}}$ is required to be evaluated at all possible values of the remaining four spins $j_{156}$, $j_{256}$, $j_{356}$, $j_{456}$. This is difficult in practice due to the potential impact on $\mathcal{Z}_{\sigma_{1-5}}$ by the complexity of the geometrical structure of $\sigma_{1-5}$, namely the existence of null directions (the real critical points resulting in some of the tetrahedra 4-D normal being null), as well as the distribution of 4-simplex volumes within $\sigma_{1-5}$ (for our fixed boundary 4-simplex, this distribution can be understood as a change of shape for each 4-simplex). Therefore, for the final goal of computing the total integral to be achieved, it is crucial to first understand the impact on spinfoam amplitude from the geometrical structure of all possible $\sigma_{1-5}$ with the same boundary 4-simplex, differing only by the position of the internal vertex $P_6$. The coordinates of $P_6$ can be expressed as:
\begin{equation}
    P_{6}=\sum_{i=1}^5 a_i\,P_i, \quad 0\leq a_i\leq 1.
\end{equation}
$P_i$ is the coordinate of the $i$-th vertex in (\ref{coordinates}). Then, we can obtain randomized samples that are evenly distributed inside the bulk region by simply randomizing $a_i$ under the restriction that the triangle inequalities are satisfied at all times. 

Additionally, in our case, we restrict the samples to satisfy the condition that for each sample, $\sigma_{1-5}$ will contain only space-like tetrahedra, i.e., all the 4-dimensional normal of tetrahedra will be time-like. In practice, out of all 20 million samples generated, only 565,247 satisfy this condition. In this work, we make this selection because we focus on the EPRL spin foam model. Our algorithm can also be generalized to Conrady-Hnybida extension, which includes the time-like tetrahedra.

In the following parts of this section, we will first discuss the results obtained from samples collected along a fixed line, which demonstrate some of our initial findings. Then, we will study the impact of the distribution of 4-simplex volumes in $\sigma_{1-5}$. Finally, we will study the results we obtained by studying samples having at least one tetrahedron 4-D normal close to the null direction, which will help further reveal the connection between geometrical structure and the corresponding spin-foam amplitude.

\subsection{Results from samples moving along fixed direction}

We begin by probing the distribution of $\mathcal{Z}_{\sigma_{1-5}}$ in terms of the position of $P_6$ inside the bulk region. The initial samples are obtained by the following procedures: First, we select point $P^{(0)}_6=(-0.0637185,-0.484167, -1.44614, -1.93792)$ as the initial data. Then, as is illustrated in FIG. \ref{15move2}, we link this point to $P_2$ in (\ref{coordinates}) and evenly sample 50 points along the edge $l_{P^{(0)}_6P_2}$, with $P^{(0)}_6=P^{(0)}_6$ and $P^{(n)}_6=P^{(0)}_6-\frac{2n-1}{100}(P^{(0)}_6-P_2), n\in\{1,\dots,50\}$. Additionally, we discover the fact that the tetrahedron normal of $(v,e)=(1,3)$ is close to null along the extension line of $l_{P_2P^{(0)}_6}$ near $P^{(0)}_6$. Therefore, we further sample 33 points near $P^{(0)}_6$ along the extension line (the dashed blue line in FIG. \ref{15move2}) of $l_{P_2P^{(0)}_6}$ to study whether there is an effect on the spin-foam amplitudes from the tetrahedron normal close to null. These additional points are labeled as $P^{(m)}_6, m\in\{-1,\dots,-33\}$ with $P^{(-1)}_6$ being closest to $P^{(0)}_6$ and
$P^{(-33)}_6=P^{(0)}+0.004408(P^{(0)}_6-P_2)$.

\begin{figure}[H]
 \centering         
 \includegraphics[width=0.3\textwidth]{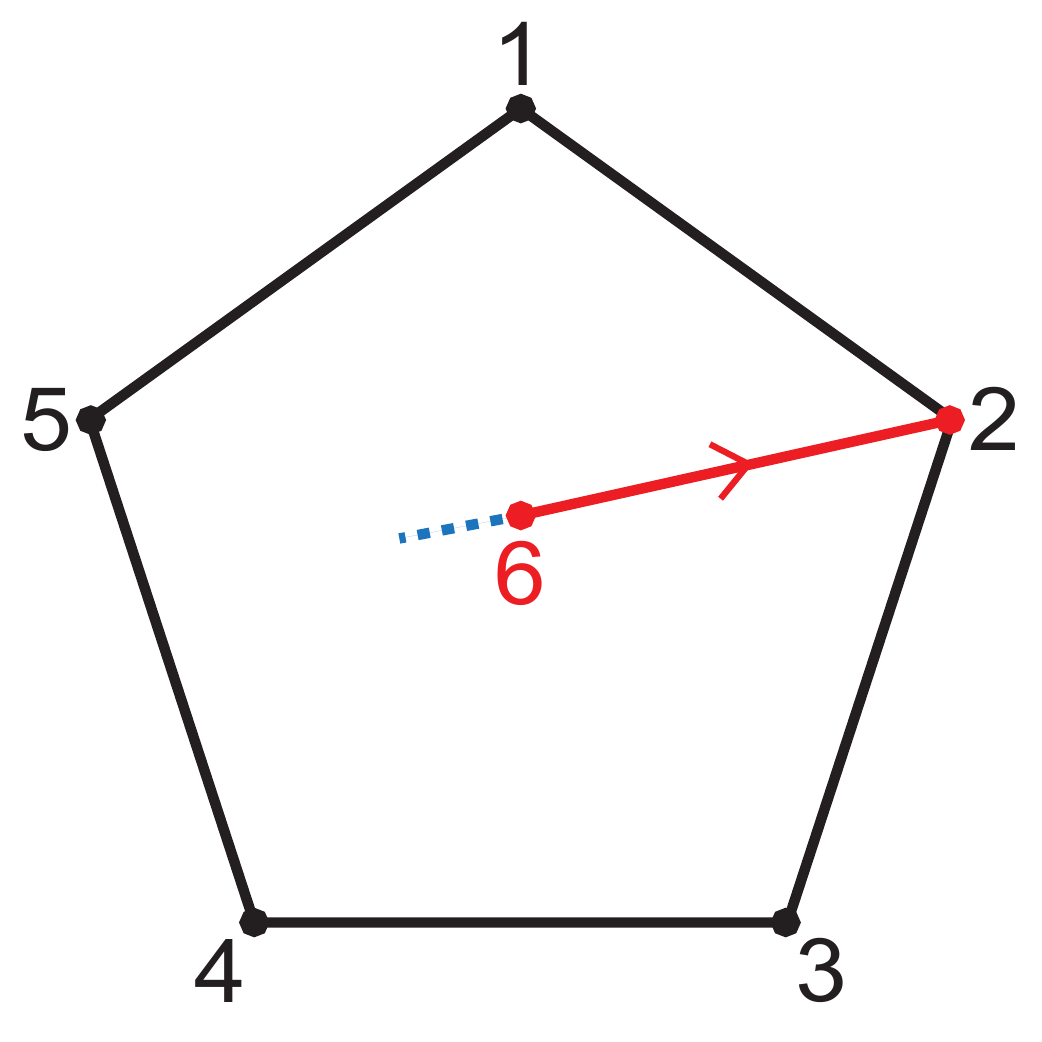}
 \caption{The red line and the dashed blue line illustrate the samples considered (see TABLE \ref{table111} for details). The arrow indicates the direction in which the samples are indexed from -33 (source of the arrow) to 50 (target of the arrow), with index 0 points to the point $P^{(0)}_6=(-0.0637185, -0.484167, -1.44614, -1.93792)$.}\label{15move2}
 \end{figure}

Since every $\sigma_{1-5}$ consists of five 4-simplices, each 4-simplex has its own 4-volume, and the sum of all five 4-simplex volumes equals the total 4-volume of $\sigma_{1-5}$, i.e., the volume of the boundary 4-simplex, it is intuitive to simply use the standard deviation of 4-simplex volumes to probe the distribution of 4-simplex volumes of $\sigma_{1-5}$:
\begin{equation}
    SD_V=\sqrt{\frac{\sum\limits^5_{i=1}(V_i-\bar{V})}{5}}
\end{equation}
where $V_i$ is the volume of the i-th 4-simplex and $\bar{V}=\frac{V_{\sigma_{1-5}}}{5}$ is the average volume of the five 4-simplices, $V_{\sigma_{1-5}}$ stands for the total volume of $\sigma_{1-5}$. It turns out that when $P_6=P_{6}^{(0)}$, the five 4-simplex volumes have the lowest standard deviation among all 20 million samples we initially generated. 

The detailed results of several key samples are shown in TABLE \ref{table111}, including the norm of leading order contribution to the partial spinfoam amplitude $|\mathcal{Z}^0_{\sigma_{1-5}}|$, the next-to-leading/leading order ratio $|\mathcal{Z}^1_{\sigma_{1-5}}/\mathcal{Z}^0_{\sigma_{1-5}}|$ and partial spinfoam amplitude up to next-to-leading order corrections $|\mathcal{Z}_{\sigma_{1-5}}|$ for each sample. In this table, the position of $P^{(n)}_6$ is labeled using their relative position to $P^{(0)}_6$ and $P_2$, with $P^{(0)}_6$ being at position 0 and $P_2$ position 1. $|t|$ and $|\vec{x}|$ are the norm of time and spatial components of the normalized normal of tetrahedron $(v,e)=(1,3)$, where the normal is close to the null direction when $|t|/|\vec{x}|\approx 1$. As presented in this table, $|\mathcal{Z}^0_{\sigma_{1-5}}|$ is large only when $SD_V$ is large, while both $|\mathcal{Z}^1_{\sigma_{1-5}}/\mathcal{Z}^0_{\sigma_{1-5}}|$ and $|\mathcal{Z}_{\sigma_{1-5}}|$ are large when either the normal of $(v,e)=(1,3)$ close-to-null or $SD_V$ is large. This result shows that the impacts to spinfoam amplitudes from close to null normal and the distribution of 4-simplex volumes are different. We further provide a more thorough analysis of this behavior below:

\begin{table}[H]
\centering
\renewcommand\arraystretch{1.3}
\begin{tabular}{|c|c|c|c|c|c|c|c|} 
 \hline
 Index & Position & $|\mathcal{Z}_{\sigma_{1-5}}|$ & $|\mathcal{Z}^0_{\sigma_{1-5}}|$ & $|\mathcal{Z}^1_{\sigma_{1-5}}/\mathcal{Z}^0_{\sigma_{1-5}}|$ & $|t|$ & $|\vec{x}|$ & $SD_V$ \\ [0.5ex] 
 \hline\hline
$-33$ & $-0.004408$ & $1.414\times 10^{-211}$ & $3.211\times 10^{-220}$ & $4.402\times 10^8$ & $505.463$ & $505.462$ & $0.0303$ \\
\hline
$\cdots$ & $\cdots$ & $\cdots$ & $\cdots$ & $\cdots$ & $\cdots$ & $\cdots$ & $\cdots$\\
\hline
$-1$ & $-0.0003$ & $8.229\times 10^{-216}$ & $3.075\times 10^{-218}$ & $268.623$ & $5.506$ & $5.415$ & $0.0305$ \\
\hline
$0$ & $0$ & $9.369\times 10^{-216}$ & $3.332\times 10^{-218}$ & $282.2$ & $5.117$ & $5.018$ & $0.0305$ \\
\hline
$1$ & $0.01$ & $2.216\times 10^{-215}$ & $6.190\times 10^{-218}$ & $359.061$ & $3.035$ & $2.865$ & $0.0319$ \\
\hline
$2$ & $0.03$ & $4.681\times 10^{-215}$ & $1.086\times 10^{-217}$ & $432.015$ & $2.092$ & $1.838$ & $0.0322$ \\
\hline
$\cdots$ & $\cdots$ & $\cdots$ & $\cdots$ & $\cdots$ & $\cdots$ & $\cdots$ & $\cdots$\\
\hline
$50$ & $0.99$ & $2.088\times 10^{-202}$ & $5.230\times 10^{-210}$ & $3.993\times 10^7$ & $1.066$ & $0.370$ & $0.1102$ \\
\hline
\end{tabular}
\caption{Detailed information on samples collected, where the relative position is computed by setting $P^{(0)}_6$ as origin which has relative position 0 and the position of $P_2$ as 1. $|t|$ and $|\vec{x}|$ are the norm of time and spatial components of the normalized tetrahedron normal of the tetrahedron $(1,3)$ (i.e. $\eta_{ab}n^an^b=-1, n^a=(t,x_1,x_2,x_3)$)}\label{table111}
\end{table}

FIG. \ref{ExpE2} (A1)-(A3) show the distribution of $|\mathcal{Z}^0_{\sigma_{1-5}}|$, $|\mathcal{Z}^1_{\sigma_{1-5}}/\mathcal{Z}^0_{\sigma_{1-5}}|$ and $|\mathcal{Z}_{\sigma_{1-5}}|$ with respect to the relative position of the aforementioned 84 samples. For all of the computations, we set $\lambda=1000$ and $\gamma=0.1$. Moreover, we have double-checked the accuracy of our computation by obtaining the exact same results with respect to multiple precision settings. The horizontal axis represents the relative position of each sample, where position 0 points to $P^{(0)}_6$ and 1 points to $P_2$. As the sample gets closer to $P_2$, both $|\mathcal{Z}^0_{\sigma_{1-5}}|$ and $|\mathcal{Z}^1_{\sigma_{1-5}}/\mathcal{Z}^0_{\sigma_{1-5}}|$ increase continuously. FIG. \ref{ExpE2} (B1)-(B2) show the close-up image of $|\mathcal{Z}^0_{\sigma_{1-5}}|$ and $|\mathcal{Z}_{\sigma_{1-5}}|$, where we reset the origin to position $-0.004408$, where the tetrahedron $(1,4)$ becomes null, such that a Log-Log graph can be plotted in the vicinity of $P^{(0)}_6$, with the rightmost data point of both graphs coming from $P^{(0)}_6$. It is clear in FIG. \ref{ExpE2} (B1)-(B2) that $|\mathcal{Z}^0_{\sigma_{1-5}}|$ decreases while $|\mathcal{Z}_{\sigma_{1-5}}|$ increases as one of the tetrahedra normal close to being null. After looking carefully at the data, we discover that the decrease of $|\mathcal{Z}^0_{\sigma_{1-5}}|$ corresponds to an increase in the determinant of the Hessian matrix, while the increase of $|\mathcal{Z}_{\sigma_{1-5}}|$ results from both the Hessian matrix as well as the measure. The orange line is the linearly fitted result where for (B1) $\mathrm{lg}(|\mathcal{Z}^0_{\sigma_{1-5}}|)=-216.329 + 0.50008 \mathrm{Log}_{10}\mathrm{lg(Pos)}$ and for (B2) $\mathrm{lg}(|\mathcal{Z}_{\sigma_{1-5}}|)=-220.63 -1.55013 \mathrm{Log}_{10}\mathrm{(Pos)}$, indicating that $|\mathcal{Z}^0_{\sigma_{1-5}}|$ is polynomial in terms of the position from $P^{(-33)}_6$ as $10^{-216.329}x^{0.50008}$ and $|\mathcal{Z}_{\sigma_{1-5}}|$ is polynomial in terms of the position from $P^{(-33)}_6$ as $10^{-220.63}\frac{1}{x^{1.55013}}$, which is slightly faster than $1/x$. The volume standard deviation of all of the samples is computed and shown in \ref{ExpE2} (C), indicating their monotonous correlation. 

In summary, we can see from both TABLE \ref{table111} and FIG. \ref{ExpE2} that along a chosen sample line, both the distribution of 4-simplex volumes and close to null tetrahedron normal will have a significant impact on spinfoam amplitudes. We will further investigate the impact caused by these two factors in arbitrary $\sigma_{1-5}$ configurations sharing the same boundary 4-simplex in the two following subsections.

\begin{figure}[H]
 \centering
 \includegraphics[height=3.5cm]{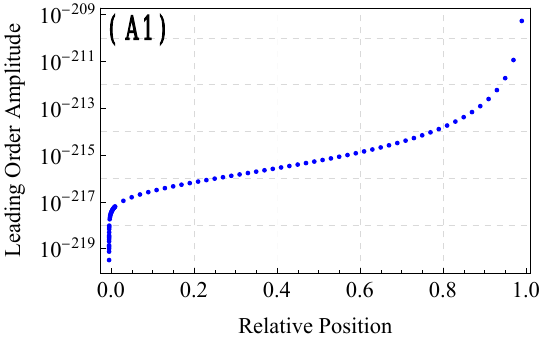}
 \includegraphics[height=3.5cm]{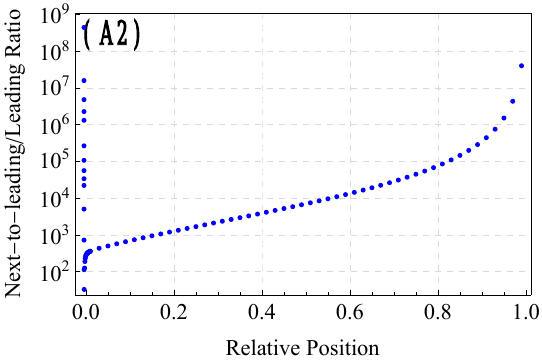}
 \includegraphics[height=3.5cm]{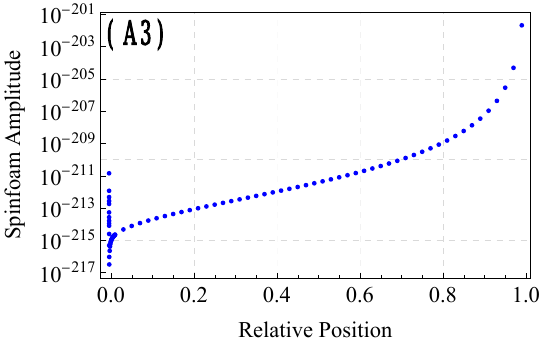}
 \includegraphics[height=3.5cm]{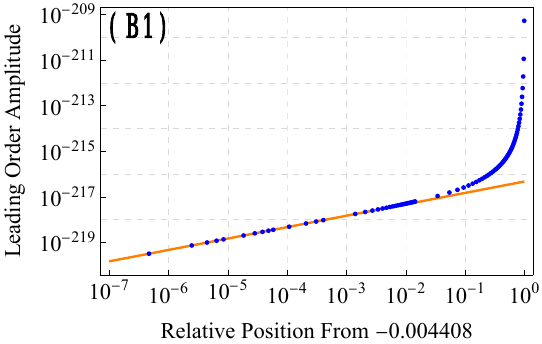}
 \includegraphics[height=3.5cm]{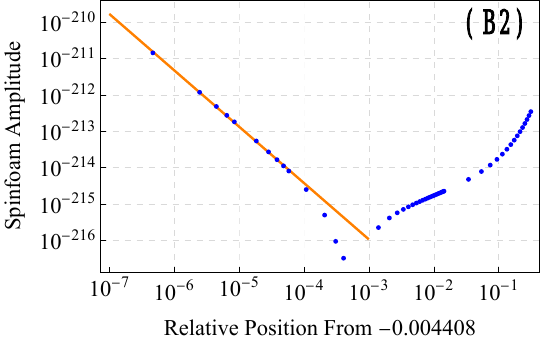}
 \includegraphics[height=3.5cm]{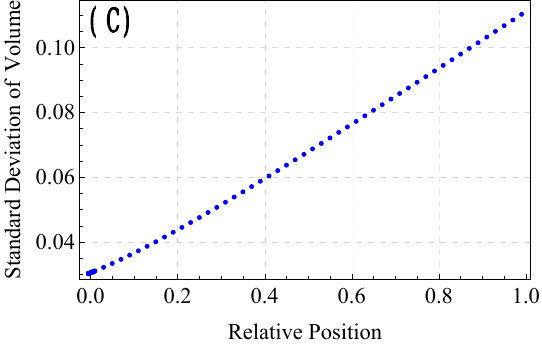}
 \caption{(A1)-(A3): spinfoam amplitudes with respect to relative positions of the internal vertex. (B1)-(B2): close-up image of $|\mathcal{Z}^0_{\sigma_{1-5}}|$ and $|\mathcal{Z}_{\sigma_{1-5}}|$ (where the logarithm of relative position is calculated by resetting the point with position $-0.0046$ as origin). The orange line is the linearly fitted result where for (B1) $\mathrm{Log}_{10}(|\mathcal{Z}^0_{\sigma_{1-5}}|)=-216.329 + 0.50008 \mathrm{Log}_{10}\mathrm{lg(Pos)}$ and for (B2) $\mathrm{Log}_{10}(|\mathcal{Z}_{\sigma_{1-5}}|)=-220.63 -1.55013 \mathrm{Log}_{10}\mathrm{(Pos)}$ (C): the standard deviation of volumes for the samples computed.}\label{ExpE2}
 \end{figure}

\subsection{The impact from distribution of 4-simplex volumes}

 In this subsection, we will study the impact on $\sigma_{1-5}$ spinfoam amplitudes from the distribution of individual 4-simplex volumes. FIG. \ref{VDis1} shows the distribution of 4-simplex volumes in terms of $SD_V$ for all $565,247$ valid samples. Since the sampling is completely random, we can conclude that for arbitrary $\sigma_{1-5}$ sharing the boundary 4-simplex $(P_1,P_2,P_3,P_4,P_5)$, their $SD_V$ values are concentrated in the neighborhood of $0.045$. 
 
 \begin{figure}[H]
 \centering         
 \includegraphics[height=5cm]{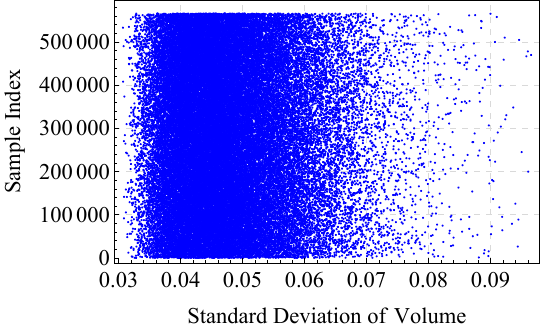}
 \includegraphics[height=5cm]{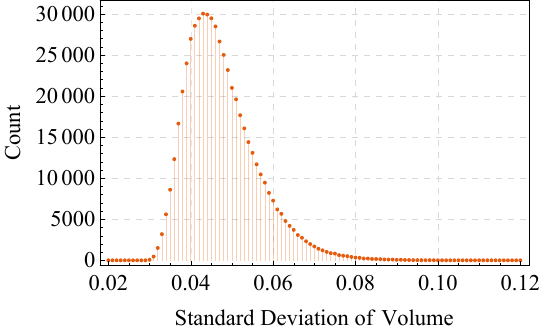}
 \caption{The distribution of all samples in terms of their standard deviation of volume. On the left side, the vertical axis keeps track of the sample index in the order of their generation. On the right side, we show a detailed plot of the density of samples with respect to the standard deviation of volume. The number of samples is counted for each $0.001$ interval from $0.02$ to $0.12$.}\label{VDis1}
 \end{figure}
 
 As can be seen from Fig. \ref{ExpE2} (C), the spinfoam amplitude in this region is small, and no divergence is observed. The distribution of $SD_V$ gets extremely sparse quickly as the standard deviation increases. This hints at the very small region where the divergence of measure might occur, thus potentially enabling the possibility to carry out the integration $A(\sigma_{1-5})$ in (\ref{int1}) numerically. However, due to the current limitation in the number of samples we can compute, we will leave the exact calculation of this extremely complicated integration for future investigation.

Due to our current computation limitation, we compute the spinfoam amplitudes from $372$ selected samples from $565,247$ valid samples, which we randomly selected such that they are evenly distributed over the entire range of $SD_V$. Also, in order to isolate the effect of tetrahedron normal on spinfoam amplitudes, we also require for every tetrahedron normal $(t,x_1,x_2,x_3)$ in each $\sigma_{1-5}$ to satisfy $\frac{|t|-|\vec{x}|}{|\vec{x}|}>0.004$, which is sufficient for minimizing the impact of tetrahedron normal and the details of which is given in the next subsection.

FIG. \ref{ExpV4} (A1), (B1), (C1) shows the impact of the standard deviation of 4-simplex volumes on the norm of $|\mathcal{Z}^0_{\sigma_{1-5}}|$, $|\mathcal{Z}^1_{\sigma_{1-5}}/\mathcal{Z}^0_{\sigma_{1-5}}|$ and $|\mathcal{Z}_{\sigma_{1-5}}|$ respectively. An overall increase can be observed for all three quantities when $SD_V$ gets larger. However, we also observe some randomness of amplitudes for samples with similar $SD_V$. We find that some samples have particularly large amplitudes when compared to other samples while sharing the same $SD_V$. To figure out the reason, we further investigated each sample in terms of its individual 4-simplex volumes as well. It turns out that, aside from the overall standard deviation, individual volumes can also have a significant impact. When we examine the individual 4-simplex volumes of each sample, we discovered that for the samples with similar $SD_V$, the sample with relatively large amplitudes also contains a very small 4-simplex volume. This can be better illustrated in TABLE \ref{table1} for the points with very similar standard deviations within a narrow range of $(0.072,0.074)$, where, for roughly the same range of $SD_V$, $|\mathcal{Z}_{\sigma_{1-5}}|$ becomes larger as the smallest 4-simplex volume gets smaller.

The relation between spinfoam amplitudes with the smallest 4-simplex volume are shown in FIG. \ref{ExpV4} (A2), (B2), (C2) for$|\mathcal{Z}^0_{\sigma_{1-5}}|$, $|\mathcal{Z}^1_{\sigma_{1-5}}/\mathcal{Z}^0_{\sigma_{1-5}}|$, and $|\mathcal{Z}_{\sigma_{1-5}}|$ respectively, showing clear correlation between the smallest 4-simplex volume and spinfoam amplitudes. \ref{ExpV4} (A3), (B3), (C3) shows the corresponding contour plot of these amplitudes with respect to both the smallest 4-volume and the standard deviation. As we can see from these three graphs, the largest amplitudes are distributed where the standard deviation of 4-simplex volumes is very large while the smallest 4-simplex volume is very small (bottom right corner of each graph).

 In summary, from the results of this subsection we can see that all spinfoam amplitudes, including$|\mathcal{Z}^0_{\sigma_{1-5}}|$, $|\mathcal{Z}^1_{\sigma_{1-5}}/\mathcal{Z}^0_{\sigma_{1-5}}|$, and $|\mathcal{Z}_{\sigma_{1-5}}|$, are hugely impacted by the distribution of 4-simplex volumes of $\sigma_{1-5}$ induced by different locations of the internal vertex. Specifically speaking, while the total volume of $\sigma_{1-5}$ remains the same, this impact mainly comes from both the smallest individual 4-simplex volume and the distribution of the five 4-simplex volumes.

 \begin{figure}[H]
 \centering
 \includegraphics[height=3.6cm]{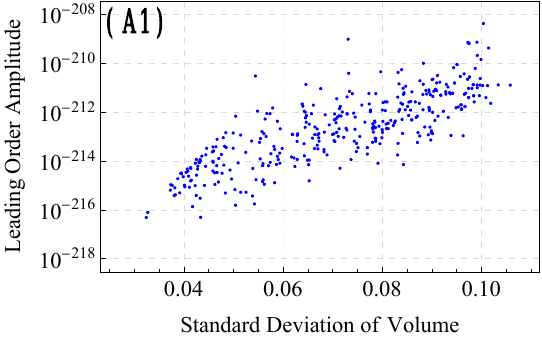}
 \includegraphics[height=3.6cm]{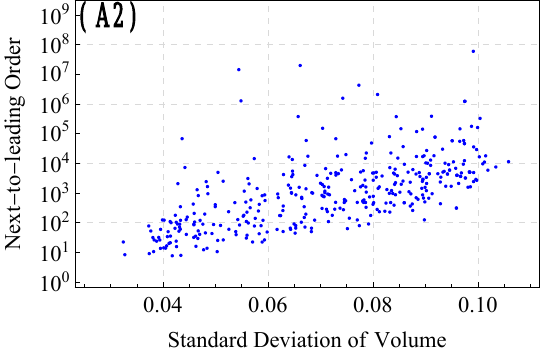}
 \includegraphics[height=3.6cm]{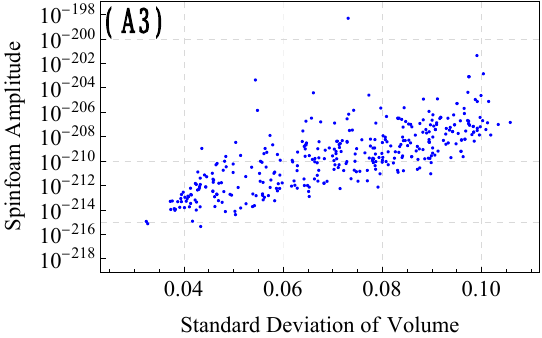}
 \includegraphics[height=3.6cm]{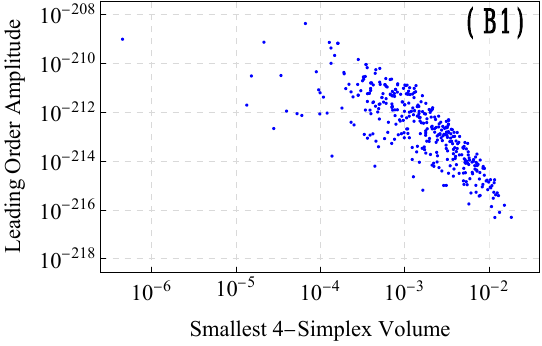}
 \includegraphics[height=3.6cm]{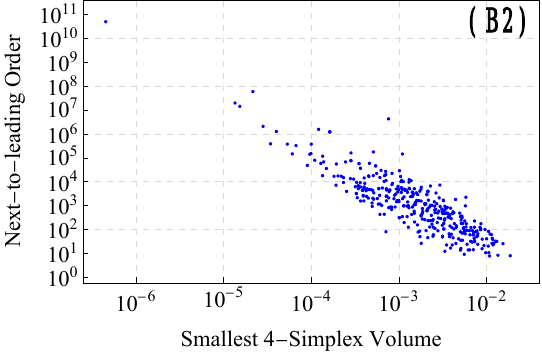}
 \includegraphics[height=3.6cm]{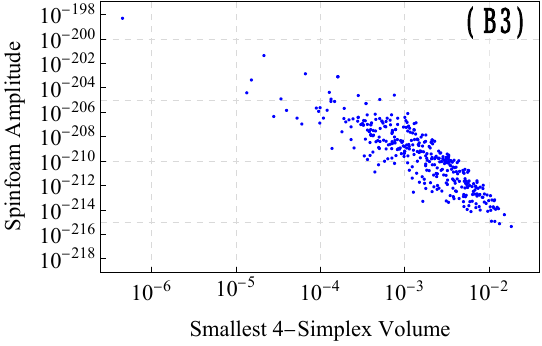}
 \includegraphics[height=5.1cm]{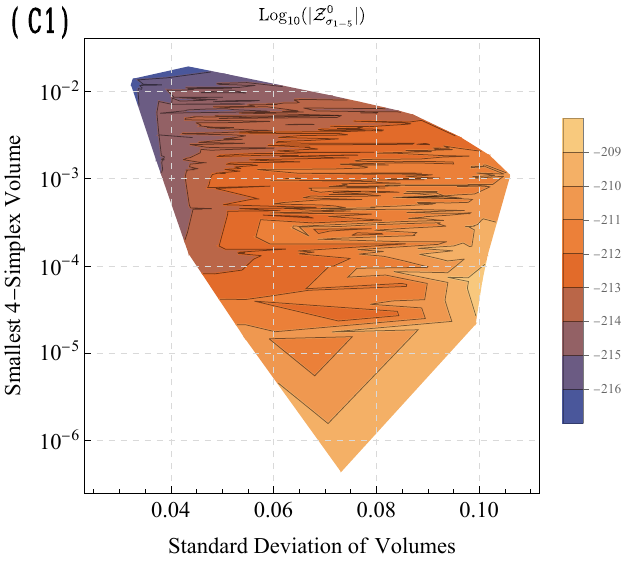}
 \includegraphics[height=5.1cm]{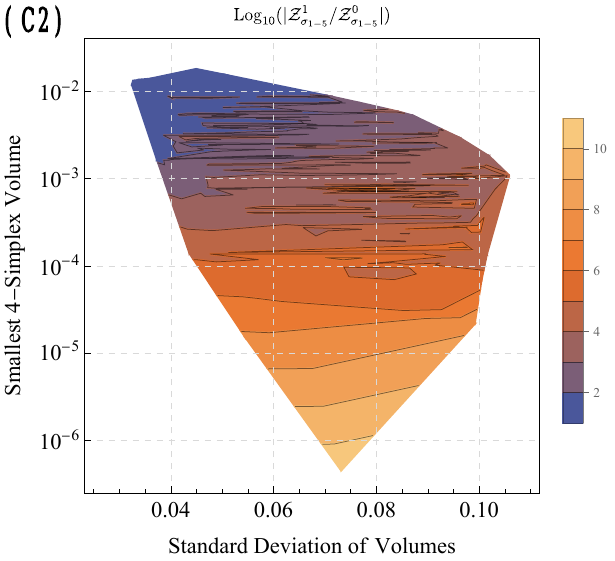}
 \includegraphics[height=5.1cm]{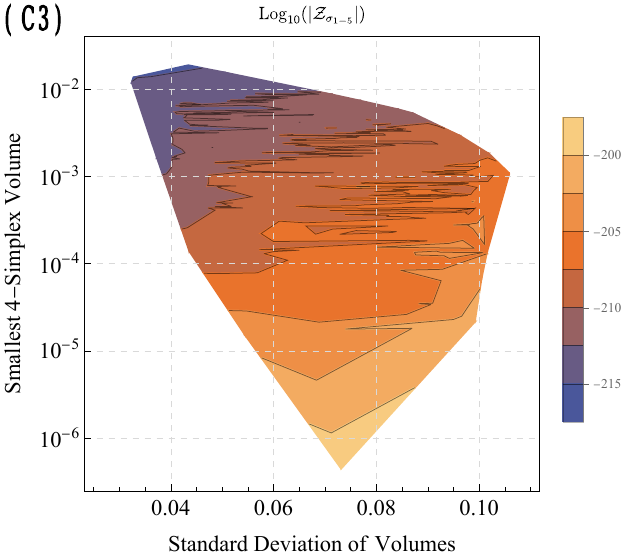}
 \caption{On the left side from top to bottom (A1), (B1), and (C1): Leading order amplitude against standard deviation, smallest 4-simplex volume, and its contour plot over both parameters. On the middle from top to bottom (A2), (B2), and (C2): Next-to-leading order correction against standard deviation, smallest 4-simplex volume, and the contour plot over both parameters. On the right side from top to bottom (A3), (B3), and (C3): the spinfoam amplitude against standard deviation, smallest 4-simplex volume, and the contour plot over both parameters.}\label{ExpV4}
 \end{figure}

\begin{table}[H]
\centering
\renewcommand\arraystretch{1.3}
\begin{tabular}{||c|c|c||c||c|c||} 
 \hline
 $|\mathcal{Z}_{\sigma_{1-5}}|$ & $|\mathcal{Z}^0_{\sigma_{1-5}}|$ & $|\mathcal{Z}^1_{\sigma_{1-5}}/\mathcal{Z}^0_{\sigma_{1-5}}|$ & $SD_V$ & smallest $V$ & \
second smallest $V$ \\
 \hline\hline
$4.773\times 10^{-199}$ & $9.597\times 10^{-210}$ & $4.974\times \
10^{10}$ & $0.0732$ & $4.539\times 10^{-7}$ & $0.000747$ \\
\hline
$6.675\times 10^{-207}$ & $3.863\times 10^{-211}$ & $17276.9$ & $0.0733$ & $0.000451$ & $0.000713$ \\
\hline
$2.771\times 10^{-207}$ & $4.175\times 10^{-212}$ & $66371$ & $0.0729$ & $0.000246$ & $0.00475$ \\
\hline
$3.100\times 10^{-208}$ & $6.991\times 10^{-212}$ & $4433.39$ & $0.0734$ & $0.000721$ & $0.00211$ \\
\hline
$3.948\times 10^{-209}$ & $8.266\times 10^{-213}$ & $4775.36$ & $0.0729$ & $0.00178$ & $0.00385$ \\
\hline
$3.389\times 10^{-209}$ & $8.589\times 10^{-213}$ & $3944.25$ & $0.0730$ & $0.00173$ & $0.00449$ \\
\hline
$8.761\times 10^{-211}$ & $1.827\times 10^{-213}$ & $478.474$ & $0.0729$ & $0.00276$ & $0.0109$ \\
\hline
$6.097\times 10^{-211}$ & $2.160\times 10^{-213}$ & $281.245$ & $0.0727$ & $0.00299$ & $0.00894$ \\
\hline
$1.620\times 10^{-211}$ & $9.212\times 10^{-214}$ & $174.842$ & $0.0723$ & $0.00324$ & $0.00819$ \\
\hline
$3.461\times 10^{-212}$ & $1.132\times 10^{-214}$ & $304.671$ & $0.0737$ & $0.00525$ & $0.0144$ \\
\hline
\end{tabular}
\caption{Comparison of a total of 10 samples with a standard deviation of 4-simplex volume within the range of $(0.072,0.074)$.}\label{table1}
\end{table}

\subsection{The impact from close to null tetrahedron normals}

From subsection B, it is suggested that as the tetrahedron normal approaches the null direction, both the next-to-leading order/leading order ratio and the total amplitude may be impacted. In this subsection, we study this effect in detail. 

Given a sample $\sigma_{1-5}$, we denote $(t_n,\vec{x}_n):=(t_n,x_n,y_n,t_n)$ as the 4-d normal of each of the 15 unique tetrahedra. To discern whether a sample is close to the null direction, we use the following standard: For each one of the fifteen unique tetrahedra contained in each sample, we use the following formula to measure how close $\vec{n}$ is to the null direction:
\begin{equation}\label{criterion}
R_n=\frac{|t_n|}{|\vec{x}_n|}-1,
\end{equation}
which is always greater than 0 and approaches 0 as $\vec{n}$ becomes close to the null direction. We select a cut-off $R_c=0.01$ as the upper bound of a normal to be considered as close to null and have the following value assigned to each tetrahedron:
\begin{equation}\label{criterion2}
    R'_n=
    \begin{cases}
        1 & \text{if } R_n\geq R_c=0.01\\
        R_n & \text{if } R_n< R_c=0.01
    \end{cases}.
\end{equation}
By taking the product of $R'_n$ of all 15 unique tetrahedra for each $\sigma_{1-5}$ sample and counting the product of $R'_n$ that is not equal to 1, we can accurately determine how close the tetrahedron normals of the arbitrary sample are to the null direction.

As a result, by applying the criterion we propose, we find that of all $565,247$ valid samples, only $21,894$, i.e., $3.87\%$ has at least one tetrahedron normal close to the null direction. Meanwhile, as we can see from FIG. \ref{NDis} (A), the distribution of samples becomes more sparse rapidly as the product of $R'_n$ decreases. In FIG. \ref{NDis} (B), it is also shown that the samples are in fact roughly evenly distributed (the count of samples dropped by two magnitudes as the $R_n'$ product becomes two magnitude smaller, indicating a linear correspondence in the range of the $R_n'$ product we analyzed) between all possible values of $R'_n$ products, making the number of points with $R'_n<10^{-6}$ very rare. Again, this hints at the small measure corresponds to the region where divergence occurs and thus enables the possibility of carrying out the integral $A(\sigma_{1-5})$ in (\ref{int1}) numerically.

\begin{figure}[H]
 \centering         
 \includegraphics[height=5cm]{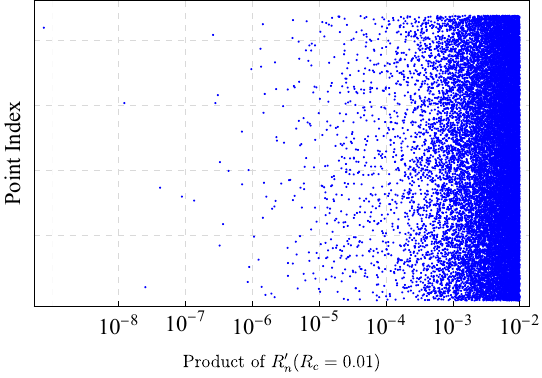}
 \includegraphics[height=5.132cm]{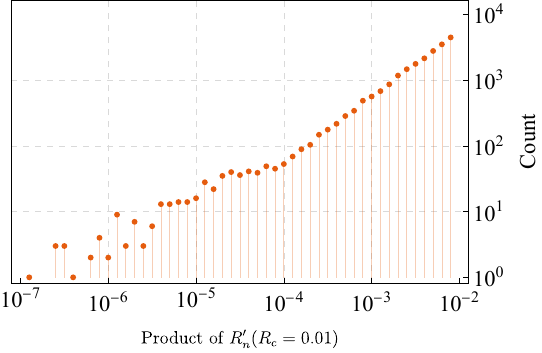}
 \caption{(Left) The logarithmic distribution of samples in terms of their individual $R'_n$ products. The blue points represent all 36331 valid samples ordered by their original generation sequence from bottom to top. (Right) The number of samples located within each $10^{1/10}$ logarithmic intervals from $10^{-7}$ to $10^{-2}$.}\label{NDis}
 \end{figure}

To study the correlation between the product of $R'_n$ and spinfoam amplitudes, we first compute $|\mathcal{Z}^0_{\sigma_{1-5}}|$, $|\mathcal{Z}^1_{\sigma_{1-5}}/\mathcal{Z}^0_{\sigma_{1-5}}|$, and $|\mathcal{Z}_{\sigma_{1-5}}|$ of 400 samples with smallest $R'_n$ products (with $R_c=0.01$), which is shown in FIG. \ref{Null2} (A1), (B1) and (C1) respectively. The horizontal axis represents the samples' $R'_n$ products, and the vertical axis represents the corresponding amplitudes.

By analyzing the samples with the largest amplitude in detail, we find that while having relatively large $R'_n$ products in general, the samples with the largest amplitudes all have a single tetrahedron normal very close to the null direction ($R'_n<0.0001$). In the meantime, the samples with the smallest spinfoam amplitudes typically have multiple normals satisfying $R'_n<0.01$ but no normals with $R'_n<0.0001$ at all. This fact suggests that the cap $R_c=0.01$ is too large for the tetrahedron normal to impact the amplitudes, resulting in the seemingly complete randomness in FIG. \ref{Null2} (A1)-(C1). Therefore, we further distinguish these samples by setting $R_c=0.0001$.

  After correlating the results obtained by using two different $R_c$ (Since $R_c=0.0001$ samples naturally satisfy $R_c=0.01$), the relation between close-to-null tetrahedron normals and the partial spinfoam amplitude is finally revealed: A single tetrahedron with its normal extremely close to the null direction greatly enlarges the next-to-leading-order/leading-order ratio $|\mathcal{Z}^1_{\sigma_{1-5}}/\mathcal{Z}^0_{\sigma_{1-5}}|$ and the total partial amplitude $|\mathcal{Z}_{\sigma_{1-5}}|$. However, the effect on the leading order amplitude $|\mathcal{Z}^0_{\sigma_{1-5}}|$ is still minimal. 
  
  To show these effects clearly, in FIG. \ref{Null2} (A2), (B2) and (C2) the samples with $R<0.0001$ are identified and colored in red. Also, the data detailing some of the key results can be found in TABLE \ref{table2}, where we have split the samples into four categories: (1) Five samples with the smallest $R'_n$ product. (2) Five red samples labeled as Red(largest), whose $|\mathcal{Z}_{\sigma_{1-5}}|$ are largest among all red samples. (3) Five red circled samples labeled as Red(smallest), whose $|\mathcal{Z}_{\sigma_{1-5}}|$ are smallest among all red samples. (4) Five blue samples labeled as Blue(largest), whose $|\mathcal{Z}_{\sigma_{1-5}}|$ are largest among all blue samples.

  There are several interesting points: Firstly, as we can see from FIG. \ref{Null2} (A1) and FIG. \ref{Null2} (B1), the correlation between the leading order amplitude and $R'_n$ products is still not obvious due to the large randomness. Only an obscure decreasing trend can be observed as the product of $R'_n$ gets smaller (which confirms the results we obtained in subsection III B). This is quite different from the case of volume distribution since volume distribution can impact both the leading order and next-to-leading order amplitudes. Secondly, the next-to-leading order corrections are heavily affected, as shown in \ref{Null2} (B2). There is a clear difference of several orders of magnitudes between the results for two different $R_c$, with the next-to-leading order corrections for the $R_c=0.0001$ case being significantly large. A similar pattern appears for $|\mathcal{Z}_{\sigma_{1-5}}|$ in \ref{Null2} (C2), where the results for the $R_c=0.0001$ case are also significantly large, suggesting that the impact of null directions on  $\mathcal{Z}_{\sigma_{1-5}}$ comes mainly from the impact on the next-to-leading order quantum corrections, making it mostly a quantum effect. Moreover, As we can see clearly from \ref{Null2} (B2) and \ref{Null2} (C2), both the blue samples and the red samples show an overall growth as the total $R'_n$ product gets smaller. This again confirms what we learned initially from subsection III B.

    Further, as we can see from TABLE \ref{table2}, samples in group (1) have significantly smaller $|\mathcal{Z}_{\sigma_{1-5}}|$ comparing to the red colored samples with largest $|\mathcal{Z}_{\sigma_{1-5}}|$ in group (2). By comparing these two groups of samples, we discovered that the smallest $R'_n$ is at a similar level for both groups. The smallest 4-simplex volume of group (1) samples are typically larger. This fact can explain their small $|\mathcal{Z}_{\sigma_{1-5}}|$. Also, the group (1) samples have roughly the same $|\mathcal{Z}_{\sigma_{1-5}}|$ as group (3) samples. By combining all the above observations, it is clear that (1) The smallest $R'_n$ rather than the total $R'_n$ product is the main factor impacting the partial spinfoam amplitude. (2) When comparing samples whose smallest $R'_n$ and the total $R'_n$ product are at a similar level, the distribution of 4-simplex volumes plays an important role in separating these samples and generating randomness in their spinfoam amplitudes.

      \begin{figure}[H]
 \centering         
 \includegraphics[height=3.5cm]{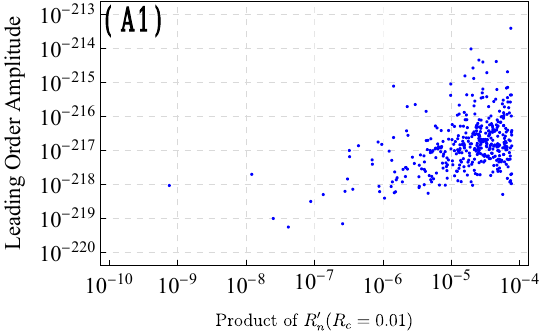}
  \includegraphics[height=3.5cm]{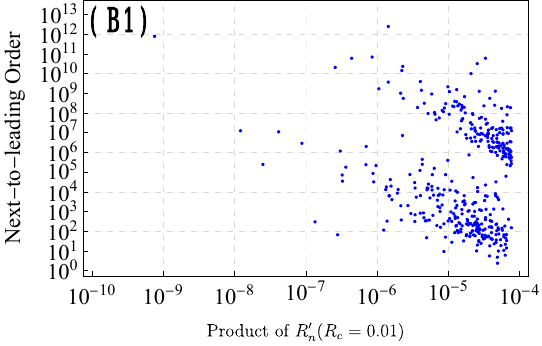}
   \includegraphics[height=3.5cm]{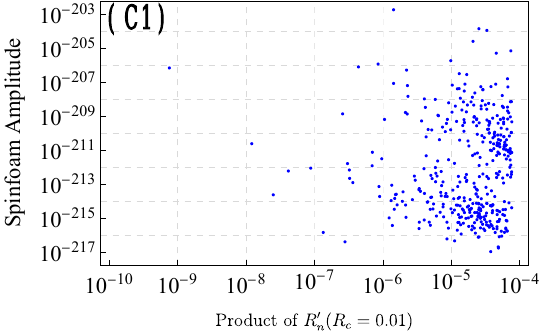}
 \includegraphics[height=3.5cm]{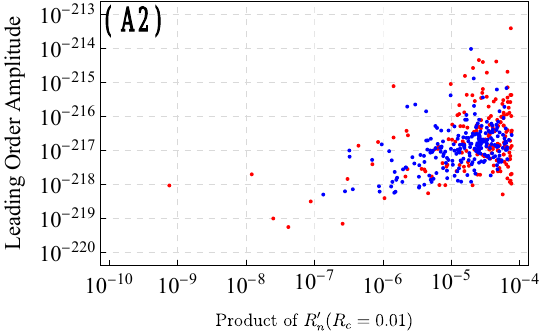}
 \includegraphics[height=3.5cm]{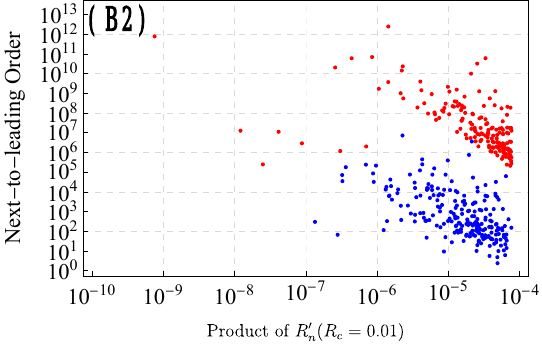}
 \includegraphics[height=3.5cm]{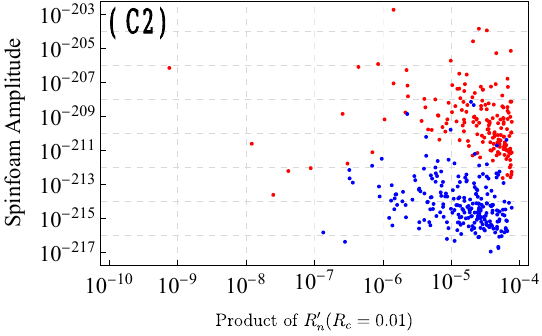}
 \caption{Top row from left to right: (A1) $|\mathcal{Z}^0_{\sigma_{1-5}}|$ when $R_c=0.01$, (B1) $|\mathcal{Z}^1_{\sigma_{1-5}}/\mathcal{Z}^0_{\sigma_{1-5}}|$ when $R_c=0.01$, (C1) $|\mathcal{Z}_{\sigma_{1-5}}|$ when $R_c=0.01$. Bottom row from left to right: (A2) the correlation between $R_c=0.01$ results (blue dots) and $R_c=0.0001$ results for $|\mathcal{Z}^0_{\sigma_{1-5}}|$ (red dots), (B2) the correlation between $R_c=0.01$ results (blue dots) and $R_c=0.0001$ results for $|\mathcal{Z}^1_{\sigma_{1-5}}/\mathcal{Z}^0_{\sigma_{1-5}}|$ (red dots), (C2) the correlation between $R_c=0.01$ results (blue dots) and $R_c=0.0001$ results for $|\mathcal{Z}_{\sigma_{1-5}}|$ (red dots).}\label{Null2}
 \end{figure}

  Also, we use FIG. \ref{NullV} to analyze how the spinfoam amplitude is affected by both the close-to-null normals and the distribution of 4-simplex volumes simultaneously. FIG. \ref{NullV} shows the distribution of $|\mathcal{Z}_{\sigma_{1-5}}|$ of 140 red-colored samples with respect to both the smallest 4-simplex volume and the standard deviation of volumes. These samples span over a select range of $R'_n$ products (from $10^{-5}$ to $10^{-4}$) in FIG. \ref{Null2}. To directly correlate the volume distribution of these samples and their spinfoam amplitudes, we need to narrow down the range of $R'_n$ products. We split the point into five groups, whose $R'_n$ products are in the range of $[10^{-5},10^{-4.8})$, $[10^{-4.8},10^{-4.6})$, $[10^{-4.6},10^{-4.4})$, $[10^{-4.4},10^{-4.2})$, $[10^{-4.2},10^{-4}]$ respectively from group (A1) to (A5).

  FIG. \ref{NullV} shows that within each group of samples of similar $R'_n$ products, the $|\mathcal{Z}_{\sigma_{1-5}}|$ is generally larger in the lower right corner of the graphs, where the standard deviation of volume is larger, and the smallest volume is smaller. It should be noted that although these samples share similar $R'_n$ products and the same order of magnitude for both the standard deviation of volume and the smallest volume, they are all randomly sampled from the bulk region. Thus, these samples can represent completely different $\sigma_{1-5}$ despite their similarities under certain criteria. Some randomness for $|\mathcal{Z}_{\sigma_{1-5}}|$ is still allowed.
  
  As we have shown in this subsection, the tetrahedron normal being close to the null direction indeed has an impact on $|\mathcal{Z}_{\sigma_{1-5}}|$. More importantly, this impact is mainly on the next-to-leading order quantum corrections, suggesting that this is mostly a quantum behavior. Also, it is confirmed in TABLE \ref{table2} that the smallest $R'_n$, rather than the smallness of the product of all $R'_n$, is the main source that causes the large partial spinfoam amplitudes.

In summary, we have shown that the spinfoam amplitudes are impacted by both how close the tetrahedron normals are to the null region and how the 4-simplex volumes are distributed. It is clear that when only a single factor is considered, the samples have noticeable randomness despite the overall trend. However, upon further looking into these randomnesses by simultaneously taking into account multiple factors and cross-analyzing all of the numerical data, we are able to fully correlate $|\mathcal{Z}_{\sigma_{1-5}}|$ with whether the sample includes some tetrahedron normals that are extremely close to the null direction, has a large standard deviation of 4-simplex volumes, or contains a very small 4-simplex volume. This result further supports the conclusion that these factors are the primary contributors to the distribution of $\sigma_{1-5}$ spinfoam amplitudes when the boundary 4-simplex is fixed.

\begin{table}[H]
\centering
\renewcommand\arraystretch{1.3}
\begin{tabular}{||c|c|c|c|c|c|c|c||} 
 \hline
 Category & $|\mathcal{Z}_{\sigma_{1-5}}|$ & $R'_n$ product & smallest $R'_n$ & $2^{\mathrm{nd}}$ smallest $R'_n$ & $SD_V$ & smallest V & $2^{\mathrm{nd}}$ smallest V \\ [0.5ex] 
 \hline\hline
$\prod R'_n<10^{-7}$ & $7 .010\times 10^{-207}$ & $7.604\times 10^{-10}$ & $9.695\times 10^{-8}$ & $0.00784$ & $0.0544$ & $0.00706$ & $0.0129$ \\
\hline
$\prod R'_n<10^{-7}$ & $2.372\times 10^{-211}$ & $1.219\times 10^{-8}$ & $9.873\times 10^{-6}$ & $0.00123$ & $0.0525$ & $0.00286$ & $0.00663$ \\
\hline
$\prod R'_n<10^{-7}$ & $2.373\times 10^{-214}$ & $2.512\times 10^{-8}$ & $0.0000591$ & $0.000425$ & $0.0406$ & $0.0178$ & $0.0226$ \\
\hline
$\prod R'_n<10^{-7}$ & $5.997\times 10^{-213}$ & $4.175\times 10^{-8}$ & $0.0000167$ & $0.00251$ & $0.0389$ & $0.0202$ & $0.0248$ \\
\hline
$\prod R'_n<10^{-7}$ & $8.868\times 10^{-213}$ & $8.867\times 10^{-8}$ & $0.0000268$ & $0.00331$ & $0.0317$ & $0.0151$ & $0.0274$ \\
\hline
Red(largest) & $1.857\times 10^{-203}$ & $1.441\times 10^{-6}$ & $1.441\times \
10^{-6}$ & $0.313$ & $0.0596$ & $0.00314$ & $0.00571$ \\
\hline
Red(largest) & $1.425\times 10^{-204}$ & $0.0000255$ & $0.0000255$ & $0.543$ & $0.0753$ & $0.00286$ & $0.00519$ \\
\hline
Red(largest) & $1.140\times 10^{-204}$ & $0.0000333$ & $0.0000333$ & $0.279$ & $0.0514$ & $0.00211$ & $0.00382$ \\
\hline
Red(largest) & $2.578\times 10^{-205}$ & $0.0000209$ & $0.0000209$ & $0.513$ & $0.0726$ & $0.00190$ & $0.00858$ \\
\hline
Red(largest) & $7.109\times 10^{-206}$ & $0.0000746$ & $0.0000746$ & $0.598$ & $0.0610$ & $0.000147$ & $0.0121$ \\
\hline
Red(smallest) & $2.373\times 10^{-214}$ & $2.512\times 10^{-8}$ & $0.0000591$ & $0.000425$ & $0.0406$ & $0.0178$ & $0.0226$ \\
\hline
Red(smallest) & $2.320\times 10^{-213}$ & $0.0000568$ & $0.0000568$ & $0.0399$ & $0.0315$ & $0.0174$ & $0.0324$ \\
\hline
Red(smallest) & $2.612\times 10^{-213}$ & $0.0000740$ & $0.0000740$ & $0.0380$ & $0.0476$ & $0.00374$ & $0.0205$ \\
\hline
Red(smallest) & $3.491\times 10^{-213}$ & $0.0000664$ & $0.0000664$ & $0.0454$ & $0.0368$ & $0.0155$ & $0.0358$ \\
\hline
Red(smallest) & $3.804\times 10^{-213}$ & $0.0000761$ & $0.0000761$ & $0.0256$ & $0.0325$ & $0.0146$ & $0.0268$ \\
\hline
Blue(largest) & $7.041\times 10^{-209}$ & $0.0000195$ & $0.00313$ & $0.00623$ & $0.0691$ & $0.00213$ & $0.00256$ \\
\hline
Blue(largest) & $4.561\times 10^{-209}$ & $0.0000214$ & $0.00244$ & $0.00876$ & $0.0457$ & $0.00246$ & $0.00446$ \\
\hline
Blue(largest) & $1.364\times 10^{-209}$ & $2.281\times 10^{-6}$ & $0.000444$ & $0.00514$ & $0.0507$ & $0.00385$ & $0.00698$ \\
\hline
Blue(largest) & $1.603\times 10^{-210}$ & $9.844\times 10^{-6}$ & $0.00212$ & $0.00465$ & $0.0447$ & $0.00349$ & $0.00635$ \\
\hline
Blue(largest) & $6.116\times 10^{-211}$ & $4.309\times 10^{-6}$ & $0.00152$ & $0.00284$ & $0.0437$ & $0.00472$ & $0.00856$ \\
\hline
\end{tabular}
\caption{Cross comparison of $|\mathcal{Z}_{\sigma_{1-5}}|$ versus multiple factors involving how close the tetrahedron normals are to the null direction and the distribution of 4-simplex volumes for four different categories of samples.}\label{table2}
\end{table}

  \begin{figure}[H]
 \centering
 \includegraphics[height=4.5cm]{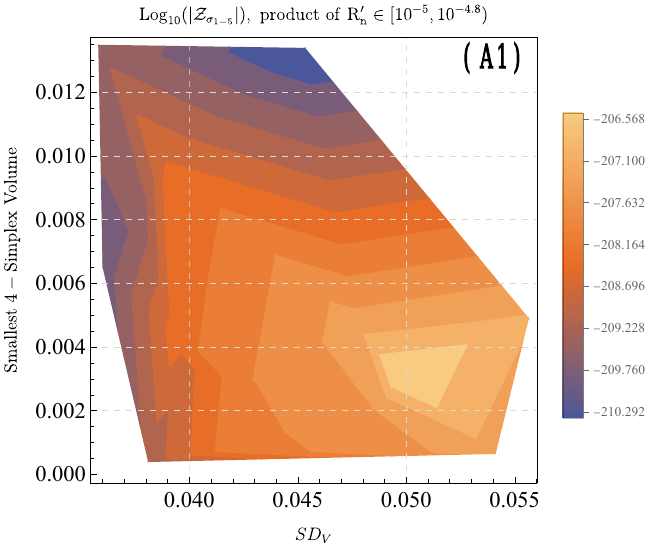}
 \includegraphics[height=4.5cm]{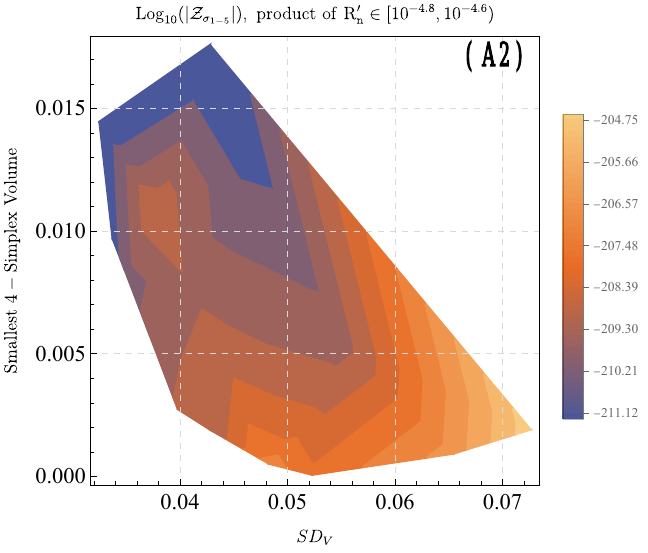}
 \includegraphics[height=4.5cm]{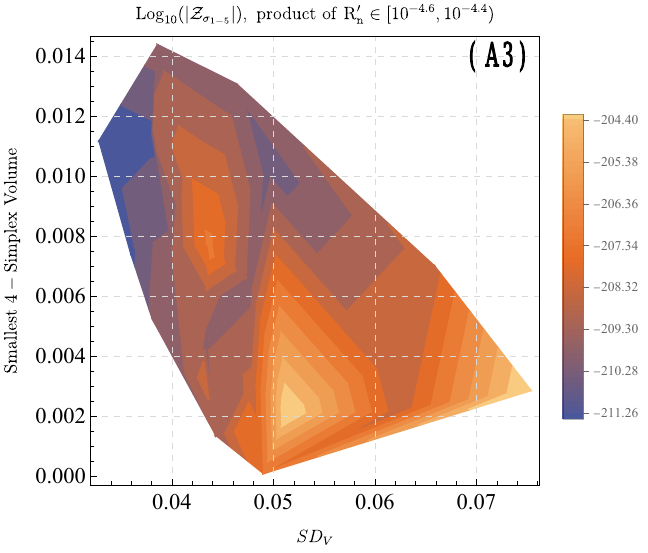}
 \includegraphics[height=4.5cm]{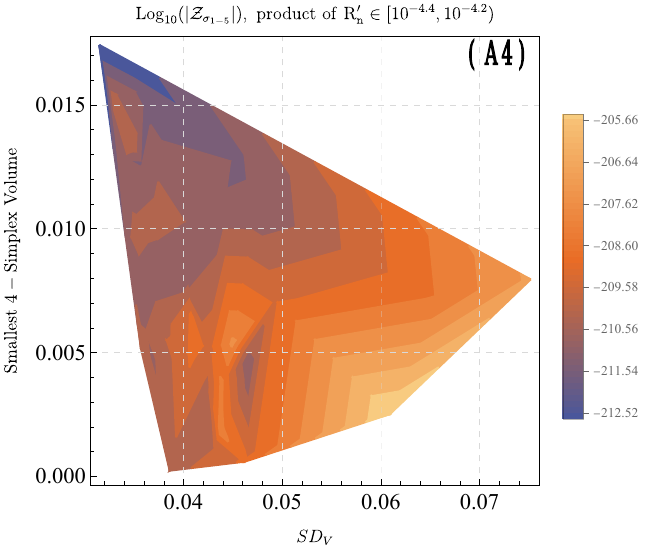}
 \includegraphics[height=4.5cm]{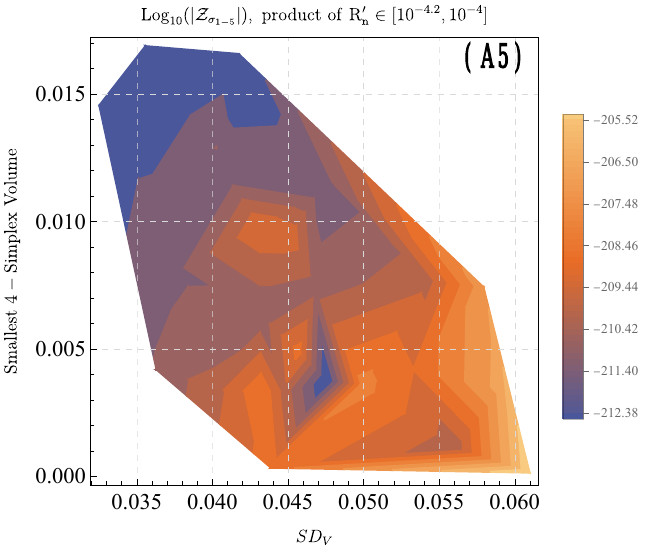}
 \includegraphics[height=4.5cm]{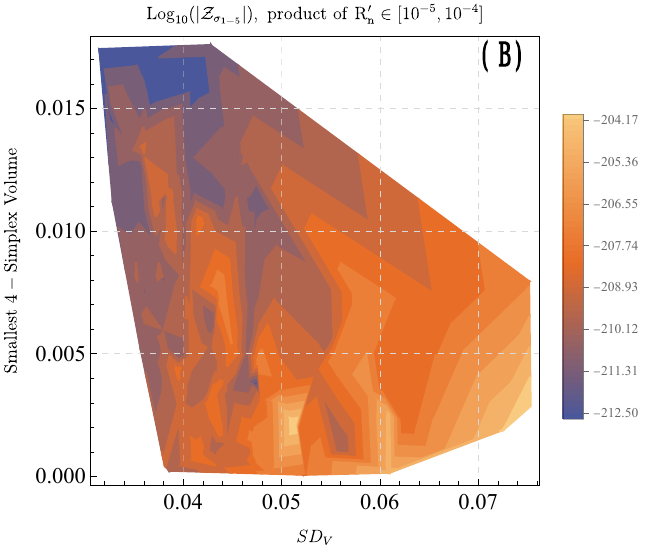}
 \caption{(A1)-(A5): The distribution of $|\mathcal{Z}_{\sigma_{1-5}}|$ within each group of samples. (B) The overall contour plot of $|\mathcal{Z}_{\sigma_{1-5}}|$ for all of the $140$ samples within the range of $R'_n\in[10^{-5},10^{-4}]$.}\label{NullV}
 \end{figure}

\section{Summary}\label{sec4}

In this paper, we discuss in detail the impact of the non-triviality of four-dimensional 1-5 Pachner move geometries on spinfoam amplitudes. We begin our numerical study by first exploring how the location of the internal vertex affects the spinfoam amplitude along a chosen line. It turns out that spinfoam amplitudes increase on both ends of the line while being relatively small in the middle. These two ends, in fact, correspond to the cases where either $\sigma_{1-5}$ has a large standard deviation of 4-simplex volumes or contains a very small 4-simplex volume, or some of the tetrahedron normal are very close to being null. We then look further into these cases using Monte-Carlo sampling and were able to fully explain the distribution of spinfoam amplitudes of all random samples obtained. There are several important findings so far:

Firstly, we use the standard deviation among individual 4-simplex volumes and the smallest 4-simplex volume to characterize the distribution of 4-simplex volumes of $\sigma_{1-5}$. As a result, strong correlation with spinfoam
amplitude is confirmed by using contour plots which take into account both the standard deviation and smallest 4-simplex volume, showing both the leading order and next-to-leading order contributions of the spinfoam amplitudes are heavily impacted by the distribution of 4-simplex volumes and the largest spinfoam amplitudes are located in general where $\sigma_{1-5}$ has the largest standard deviation and consequently the smallest, second smallest and third smallest 4-simplex volume are very small.

Secondly, we study the impact of close-to-null tetrahedron normal by introducing a cut-off criterion (\ref{criterion2}). We find that the impact of tetrahedron normal is very sensitive to the criterion we proposed, and a significant impact is only visible after we correlate the results by imposing two different cut-offs, ultimately showing that a single tetrahedron normal extremely close to the null direction is more important than several close-to-null normal who are actually less close to the null direction. Moreover, it is also revealed that the impact of close-to-null tetrahedron normal is only obvious on the next-to-leading order term, suggesting that this impact is mostly a quantum effect.

Thirdly, while strong correlations are found for both cases, the randomness of individual samples still plays an important role in determining the spinfoam amplitudes at first glimpse. However, after we carefully examine all the numerical data obtained and made cross reference using all factors simultaneously, we are able to explain the value of spinfoam amplitude of almost every individual sample we studied, suggesting that the distribution of 4-simplex volumes and close-to-null normals are indeed the most important factors in determining spinfoam amplitude. Moreover, based on the definition given in (\ref{action}), we believe this conclusion is true for all graphs when considering spinfoam large-$j$ asymptotics.

 Despite the many discoveries we made in this paper, we are still unable to compute the full amplitude (\ref{int1}). This is because the values of spinfoam amplitudes vary dramatically, causing a Monte Carlo integration to result in an error of the same order of magnitude as the largest amplitude. This problem will be addressed in the future in the following two steps: First, we will rewrite our program completely in more efficient languages such as Python or Julia. Since our algorithm involves intensive access to large matrices over billions of entries, by utilizing the fully vectorized operations and much more efficient cycle execution provided by these languages we can drastically reduce computation time and expect to have a sample set of several orders of magnitude larger than the current one, which only consists of about 1000 samples in total. Second, since singular directions are crucial in the computation, we will try to incorporate other advanced integration methods, such as the Markov chain Monte-Carlo method, which excels at computing the contribution of divergent samples in the integral region when compared to the ordinary Monte-Carlo method. After taking these two steps, we will then move on to fully understand $\sigma_{1-5}$ spinfoam amplitude and its relation with the spinfoam amplitude of its boundary 4-simplex. Also, the finiteness of 1-5 move including the timelike tetrahedra in the large-$j$ limit can be computed using our algorithm to compare the results in \cite{Borissova:2024txs}}.

\section*{Acknowledgments}

MH receives support from the National Science Foundation through grants PHY-2207763 and the College of Science Research Fellowship at Florida Atlantic University. MH, HL are supported by research grants provided by the Blaumann Foundation. The Government of Canada supports research at Perimeter Institute through Industry Canada and by the Province of Ontario through the Ministry of Economic Development and Innovation. This work benefits from the visitor's support from Beijing Normal University, FAU Erlangen-N\"urnberg, the University of Western Ontario, and Perimeter Institute for Theoretical Physics.

\bibliographystyle{unsrt}
	\bibliography{sample}

\end{document}